\documentclass[10pt,a4paper]{article}
\usepackage{float}
\usepackage{jheppub_kim}
\usepackage{epstopdf}
\usepackage{pdflscape}
\usepackage{amsmath}
\usepackage{amssymb}
\usepackage{dcolumn}
\usepackage{bm}
\usepackage{color}
\usepackage{amsmath}
\usepackage{breqn}
\usepackage{booktabs}
\usepackage{graphicx}
\usepackage{epsfig}
\usepackage{caption,subcaption}
\usepackage{amsfonts}
\usepackage{graphicx}
\usepackage{parskip}
\usepackage{xcolor} % add this in preamble
\usepackage{dcolumn}
\UseRawInputEncoding

\begin{document}

\title{Ricci-Cubic Holographic Dark Energy: Confronting Observations, Stability and the Cosmic Coincidence Problem}

\author[a]{Aritra Sanyal}
\author[b]{Prabir Rudra,}

\affiliation[a] {Department of Mathematics, Jadavpur University, Kolkata-700 032, India.}

\affiliation[b] {Department of Mathematics, Asutosh College,
Kolkata-700 026, India.}

\emailAdd{aritrasanyal1@gmail.com}
\emailAdd{prudra.math@gmail.com}

\abstract{
In this work, we constrain the parameter space of the Ricci-Cubic Holographic Dark Energy (RCHDE) model using several observational datasets, including Hubble parameter measurements, cosmic chronometer (CC) data, Baryon Acoustic Oscillation (BAO) data, and recent DESI observations. The RCHDE model is constructed from a cubic curvature invariant formed through cubic contractions of the Ricci and Riemann tensors. To estimate the model parameters, we employ the Markov Chain Monte Carlo (MCMC) sampling technique within a Bayesian inference framework. The resulting likelihood contours provide both marginalized and joint posterior distributions of the model parameters. The best-fit cosmological evolution predicted by the RCHDE model is reconstructed and compared with observational $H(z)$ measurements as well as with the standard $\Lambda$CDM cosmological model. The best-fit value obtained in our model exhibits a moderate Hubble tension of approximately $2.3\sigma$ with respect to the reference value for $\Lambda$CDM. While this indicates a noticeable discrepancy, it remains significantly lower than the $\sim 5\sigma$ tension typically reported between early- and late-Universe measurements, suggesting a partial alleviation of the tension. In addition to the statistical parameter estimation, we perform an enhanced machine learning analysis using observational Hubble parameter data. Several supervised regression algorithms are implemented to reconstruct the expansion history of the universe and to test the predictive capability of the RCHDE model from a data-driven perspective. Various graphical analyses are presented to illustrate the performance of the machine learning models. The results demonstrate a strong consistency between the predictions of the machine learning models, the theoretical RCHDE model, and the observational data. We have done a comparative stability analysis between different holographic dark energy models using the squared speed of sound, where it is seen that the RCHDE model does not have any upper hand over its counterparts. Finally, the cosmic coincidence problem is tested to compare the efficiency of the RCHDE model in comparison to other models. It is found that the RCHDE model produced a significant alleviation to the cosmic coincidence problem, outshining its counterparts.}

\maketitle

\section{Introduction}

The universe has recently entered an accelerated expansion phase, as shown by observations of SN Ia supernovae (S. Perlmutter et al. 1999; A. G. Riess et al. 1998). The era that was ruled by matter for a long time has clearly ended. There are other theoretical frameworks that can explain this strange phenomenon, but the cosmological constant $\Lambda$ is the most commonly used one (P. J. E. Peebles \& B. Ratra (2003)). However, to solve the dynamical nature, more degrees of freedom beyond the usual framework of particle physics and general relativity (A. Einstein 1916) must be included. Furthermore, it is never easy to explain the universe's whole thermal history, including early inflation. Two distinct methods can be used to modify general relativity. Modified gravity theories (S. Nojiri, S. D. Odintsov \& V. K. Oikonomou 2017; S. Nojiri \& S. D. Odintsov (2007); S. Capozziello et al. 2019) were originally created by adding changes in the geometrical sector. Altering the matter sector and adding exotic, negative-pressure components—known as dark energy (P. Brax 2018)—is the alternative. As desired, both of these ideas make use of additional degrees of freedom.

The literature contains a variety of dark energy candidates. Notable examples include scalar field models and Chaplygin gas models (A. Kamenshchik et al. 2001; V. Gorini et al. 2003; P. Rudra et al. 2012; R. Chowdhury \& P. Rudra (2013); P. Rudra 2012). According to the holographic principle, which derives from black hole thermodynamics, a system's entropy is determined by its area rather than its volume (L. Susskind 1995; R. Bousso 2002). This holographic principle, which is also related to string theory (L. Susskind 1995; G't Hooft 1993), has led to the development of holographic dark energy (HDE) (S. D. H. Hsu 2004; R. Horvat 2004; M. Li 2004). The greatest distance that may be covered by a quantum field theory is known to have connections with an ultraviolet cutoff (A. G. Cohen, D. B. Kaplan, \& A. E. Nelson 1999). The vacuum energy, which is a type of dark energy with a holographic origin, is directly related to this ultraviolet cutoff. For a comprehensive overview of HDE, see (S. Wang, Y. Wang, \& M. Li 2017). Both the basic and expanded versions of HDE have been the subject of much research, and the model has proven to be highly effective over time (R. Horvat 2004; Q. G. Huang \& M. Li (2004); D. Pavon \& W. Zimdahl (2005); Y. G. Gong 2004; E. N. Saridakis 2008; M. R. Setare \& E. C. Vagenas (2008); S. Maity \& P. Rudra (2022)). The adaptability of HDE models with observational data has been one of their main advantages (X. Zhang \& F. Q. Wu (2005); M. Li et al. 2009; J. Lu et al. 2010).

The HDE density is known to be proportional to the inverse square of the infrared cutoff $L$, which is determined by
\begin{equation}\label{hdedensity}
\rho_{DE} = \frac{3c}{\kappa^{2}L^{2}}
\end{equation}
where $\kappa$ is the gravitational constant and $c$ is a parameter. However, there is no consensus on the appropriate infrared cutoff for the cosmic application of the holographic principle. The most popular options are the particle horizon and the Hubble radius, but neither of them can propel the cosmic acceleration (S. D. H. Hsu 2004). Eventually, it is the future event horizon that is the most appropriate cutoff for the situation (M. Li 2004). Despite being a good fit for the situation, this decision has certain logical issues. The future evolution of the dark energy density actually determines its present value, which is a difficult idea to comprehend. Therefore, more efforts have been made to identify modified holographic dark energy models in which the DE density is dependent on past and present evolution rather than future evolution. Agegraphic dark energy models are those in which the infrared cutoff can be determined by quantities based on the universe's historical characteristics. Here, the age of the universe or the conformal time can be used to determine the infrared cutoff (R. G. Cai 2007; H. Wei \& R. G. Cai (2008); M. Jamil \& E. N. Saridakis (2010)). If dark energy originates from fundamental spacetime physics, then it is natural that its energy density should be constructed from the geometric invariants (quantities that do not change under coordinate transformations) of spacetime curvature. Since these invariants remain the same for all observers, they represent true physical properties of spacetime. Moreover, curvature invariants have a natural connection with cosmological dynamics and are compatible with quantum gravity ideas. This is because the geometry of spacetime has a direct connection with horizon entropy, which can further be connected to the energy bounds of the system. In literature, there exist HDE models that use geometrical invariants as the IR cutoff. Use of geometrical invariants as an IR cutoff is a sound concept since the evolution depends on the present and local features of the universe. The most common model is the Ricci Holographic dark energy (RHDE) (Gao et al. 2009), where the inverse square root of the Ricci scalar is used as the IR cutoff. RHDE has interesting cosmological applications widely available in literature (Gao et al. 2009; C. J. Feng \& X. J. Li (2009); M. Suwa \& T. Nihei (2010); S. Nojiri \& S. D. Odintsov (2017); S. Nojiri, S. D. Odintsov, E. N. Saridakis 2019). Some phenomenological issues with minimal HDE can be found in (E. O. Colgain \& M. M. Sheikh-Jabbari (2021); A. Banerjee et al. 2021). In spite of the above advantages, the RHDE model suffers from a handful of drawbacks. It was found that the model showed classical instability ($c_s^2<0$) in perturbations of dark energy, which can cause growth of unphysical fluctuations (S. del Campo, J. C. Fabris, R. Herrera, W. Zimdahl, Phys. Rev. D 87, 123002 (2013);  K. Karwan, T. Thitapura, JCAP 1201: 017 (2012); R. Herrera, W. S. Hipolito-Ricaldi, N. Videla, JCAP 08: 065 (2016)). Moreover, there are some reviews that also discuss this problem in detail (M. Li, S. Wang, Y. Wang, X.-D. Li, Phys. Rept. 509, 1 (2011); S. Nojiri, S. D. Odintsov, Phys. Rept. 505, 59 (2011); E. J. Copeland, M. Sami, S. Tsujikawa, Int. J. Mod. Phys. D 15, 1753 (2006)). RHDE works well in explaining the late-time acceleration, but it may produce problems when describing the early radiation-dominated era, unless additional assumptions are added (E.-K. Li, Y. Zhang, J.-L. Geng, Phys. Rev. D 90, 083534 (2014);  S. del Campo, J. C. Fabris, R. Herrera, W. Zimdahl, Phys. Rev. D 83, 123006 (2011)). Although RHDE attempts to reduce the Cosmic Coincidence Problem, the explanation is not completely satisfactory. Later studies showed that the improvement is only partial (I. Durán, D. Pavón, Phys. Rev. D 83, 023504 (2011)).

Since invariants are essential to physics, it follows that we can represent the infrared cutoff using different invariants or their combinations, creating various types of holographic dark energy in the process. Moreover, the above-discussed drawbacks of the RHDE model motivate such formulations. The Gauss-Bonnet invariant would be the most straightforward example of such an extension. To address this, Saridakis described the infrared cutoff using a combination of the Gauss-Bonnet invariant and the Ricci scalar (E. N. Saridakis 2018). This model is named Ricci-Gauss-Bonnet holographic dark energy (RGBHDE), and it displays a number of fascinating cosmological characteristics, such as the universe's whole thermal history, from radiation to matter to the dark energy-dominated epoch. From a mathematical perspective, there may be numerous options because we are discussing invariants. Their mathematical relevance is undeniable, but it is reasonable that their physical significance and importance to cosmology are separate issues that require further investigation. Two Riemann tensor contractions were used to create the Gauss-Bonnet invariant. Therefore, the spectrum of this quadratic theory matches that of Einstein gravity. But there is a serious issue with this construction coming from differential geometry. From the Gauss-Bonnet theorem of differential geometry, we see that the Gauss-Bonnet term $G=R^{2}-4R_{\mu\nu}R^{\mu\nu}+R_{\mu\nu\rho\sigma}R^{\mu\nu\rho\sigma}$ is topological (not dynamical) in 4-dimensional general relativity. When we include this term in an action, the integral of $G$ contributes only a boundary term, and so it does not change Einstein’s field equations directly. Moreover, this boundary term can easily vanish when we choose suitable boundary conditions. So the Gauss-Bonnet term has very little influence on the field equations unless it is modified or coupled. In addition to this, the model is expected to suffer from the instability problem, which was present in RHDE (A. Jawad, A. Pasqua, S. Chattopadhyay,  Astrophys Space Sci 344, 489–494 (2013)). Therefore, using $G$ directly in the energy density may lack strong justification. Motivated by this, Ricci-cubic holographic dark energy (RCHDE) was proposed by Rudra (P. Rudra 2023), using the cubic invariant $P$, which in turn is formed from all possible combinations of the cubic contractions of Riemann and Ricci tensors. The cubic gravity theory (P. Bueno \& P. A. Cano (2016)) and even generalized $f(P)$ gravity theories (C. Erices, E. Papantonopoulos \& E. N. Saridakis 2019; M. Marciu 2020; K. Giri \& P. Rudra (2022)) have been developed using the cubic invariant. In the RCHDE model proposed by Rudra, the infrared cutoff is determined by combining the cubic invariant with the Ricci scalar. One may think of this as an extension of the RHDE and RGBHDE models. The cosmological properties of RCHDE are intriguing and offer some enhanced dynamical capabilities. This is because the two curvature invariants are controlled by two model parameters, offering a richer cosmological structure.

In this paper, we further develop the RCHDE model by constraining the free parameters of the model using observational data. Using observational data to constrain a cosmological model is crucial to its scientific use for advancing our knowledge of the cosmos. Model parameters determine the specific behavior and predictions of the model. The model becomes more realistic and precise by constraining them with data (such as supernovae, cosmic microwave background, and large-scale structure, etc.). Without observational restrictions, parameters can fluctuate greatly, producing imprecise or pointless forecasts. Data increases the model's predictive power by reducing the range of probable values. The model can be used to generate accurate predictions (e.g., regarding galaxy formation, cosmic expansion, etc.) once parameters are limited. These predictions may be put to the test by subsequent observations, which could improve the model or uncover new physics. By comparing the performance of each model, constraining parameters lets us determine which model best captures reality. The paper is organized as follows: Section~2 discusses the important aspects of RCHDE. Section~3 is dedicated to a detailed observational data analysis. Section~4 presents an enhanced machine learning analysis of the model. Finally, the paper ends with a short conclusion in Section~5.

\section{Ricci-cubic Holographic dark energy}
There is a family of cubic theories with six parameters for the cubic case, and their spectrum is the same as the Einstein gravity [P. Bueno \& P. A. Cano (2016)]. In a 4-dimensional spacetime, a
general non-topological cubic term would be given by (P. Bueno \& P. A. Cano (2016); C. Erices, E. Papantonopoulos \& E. N. Saridakis (2019)),
$$P=\beta_{1}R_{\mu~~\nu}^{~~\rho~~\sigma}R_{\rho~~\sigma}^{~~\gamma~~\delta}R_{\gamma~~\delta}^{~~\mu~~\nu}
+\beta_{2}R_{\mu\nu}^{\rho\sigma}~R_{\rho\sigma}^{\gamma\delta}~R_{\gamma\delta}^{\mu\nu}+\beta_{3}R^{\sigma\gamma}R_{\mu\nu\rho\sigma}{R^{\mu\nu\rho}}_{\gamma}
+\beta_{4}RR_{\mu\nu\rho\sigma}R^{\mu\nu\rho\sigma}$$
\begin{equation}\label{cubicterm}
+\beta_{5}R_{\mu\nu\rho\sigma}R^{\mu\rho}R^{\nu\sigma}+\beta_{6}R^{\nu}_{\mu}R^{\rho}_{\nu}R^{\mu}_{\rho}+\beta_{7}R_{\mu\nu}R^{\mu\nu}R+\beta_{8}R^{3}
\end{equation}
where $\beta_{i}$ are parameters. The following parameter requirements are met if the theory's spectrum matches that of general relativity
\begin{equation}\label{cond1}
\beta_{7}=\frac{1}{12}\left(3\beta_{1}-24\beta_{2}-16\beta_{3}-48\beta_{4}-5\beta_{5}-9\beta_{6}\right)
\end{equation}
\begin{equation}\label{cond2}
\beta_{8}=\frac{1}{72}\left(-6\beta_{1}+36\beta_{2}+22\beta_{3}+64\beta_{4}+3\beta_{5}+9\beta_{6}\right)
\end{equation}

The Friedmann-Lemaitre-Robertson-Walker (FLRW) metric, which models a homogeneous and isotropic universe, is given by
\begin{equation}\label{flrwmet}
ds^{2}=-dt^{2}+a^{2}(t)\left[\frac{dr^{2}}{1-kr^{2}}+r^{2}\left(d\theta^{2}+\sin^{2}\theta d\phi^{2}\right)\right]
\end{equation}
where $a(t)$ is the cosmological scale factor and $k$ is the
spatial curvature, such that $k=-1, 0, +1$ corresponds to open,
flat and closed spatial geometry, respectively. Here we will concentrate on the flat geometry, which may be easily extended to the open and closed universes. In order to have uniform dimensions, we express the IR cutoff of RCHDE as a combination of $R$ and $P^{1/3}$ as follows (P. Rudra 2023)
\begin{equation}\label{irrc}
\frac{1}{L^{2}}=-\alpha R+\lambda P^{1/3}
\end{equation}
where the constants $\alpha$ and $\lambda$ are model parameters. It can be clearly seen that for $\lambda=0$, we retrieve the usual RHDE, while for $\alpha=0$, we get a pure cubic HDE. Using
eqn.(\ref{irrc}) in eqn.(\ref{hdedensity}) we get the energy
density of the RCHDE as,
\begin{equation}\label{denrc1}
\rho_{DE}=\frac{3}{\kappa^{2}}\left(-\alpha R+\lambda
P^{1/3}\right)
\end{equation}
where the constant $c$ has been absorbed in the model parameters
$\alpha$ and $\lambda$ for convenience. Now for the flat FLRW
geometry, the Ricci scalar and the cubic invariant are respectively
given by,
\begin{equation}\label{ricfrw}
R=-6\left(2H^{2}+\dot{H}\right)
\end{equation}
and
\begin{equation}\label{cubfrw}
P=6\tilde{\beta} H^{4}\left(2H^{2}+3\dot{H}\right)
\end{equation}
where $H=\dot{a}/a$ is the Hubble function and the dots denote
derivatives with respect to time. Moreover in the above expression
we have defined $\tilde{\beta}$ as,
\begin{equation}
\tilde{\beta}\equiv -\beta_{1}+4\beta_{2}+2\beta_{3}+8\beta_{4}
\end{equation}
It should be mentioned here that for the cubic invariant, we have
considered derivatives only up to first order, so that the FLRW
equations are of the second order. The condition for the second
order field equations are satisfied if we consider,
\begin{equation}
\beta_{6}=4\beta_{2}+2\beta_{3}+8\beta_{4}+\beta_{5}
\end{equation}
Using these, the energy density of Ricci-cubic HDE becomes,
\begin{equation}\label{denrc2}
\rho_{DE}=\frac{3}{\kappa^{2}}\left[6\alpha\left(2H^{2}+\dot{H}\right)+\lambda
\left\{6\tilde{\beta} H^{4}\left(2H^{2}+3\dot{H}\right)\right\}^{1/3}\right]
\end{equation}
The dark energy component in this model emerges effectively from the Ricci-cubic holographic corrections to the gravitational sector, and thus represents a geometrically induced fluid rather than a fundamental energy component.
%In terms of the scale factor the above expression can be written as,
%\begin{equation}\label{rhode}
%\rho_{DE}=\frac{3}{\kappa^{2}}\left[\frac{6\alpha\left(\dot{a}^{2}+a\ddot{a}\right)}{a^{2}}-\frac{6^{1/3}\lambda}{a^{2}}\left(\beta \dot{a}^{4}\left(\dot{a}^{2}-3a\ddot{a}\right)\right)^{1/3}\right]
%\end{equation}
Now the first FLRW equation is given by,
\begin{equation}\label{flrweq1}
3H^{2}=\kappa^{2}\left(\rho_{m}+\rho_{DE}\right)
\end{equation}
where $\rho_{m}$ is the energy density of matter. The equation of
state (EoS) parameter of matter is given by
$w_{m}=p_{m}/\rho_{m}$, where $p_{m}$ is the pressure of matter. Similarly the second FLRW equation is given by
\begin{equation}\label{flrweq2}
\dot{H}+H^{2}=-\frac{\kappa^2}{6}\Big[\rho_{m}+\rho_{DE}+3\left(p_{m}+p_{DE}\right)\Big]
\end{equation}
Finally, the matter sector follows the conservation relation given
by,
\begin{equation}\label{conseq}
\dot{\rho}_{m}+3H\left(\rho_{m}+p_{m}\right)=0
\end{equation}
Using eqns.(\ref{flrweq1}) and (\ref{conseq}) one can fully
determine the evolution of the universe, provided the matter
equation of state is known. Generally, we consider a pressureless
matter sector, i.e. $p_{m}=0$ leading to $w_{m}=0$. Using this in
the conservation equation (\ref{conseq}) we get the matter energy
density as $\rho_{m}=\rho_{m0}a^{-3}$, where $\rho_{m0}$ is the
density of matter in the present time. Thus, we have the matter
energy density and also from eqn.(\ref{denrc2}) we have the dark
energy density. Using them in the FLRW equation (\ref{flrweq1}) we have a differential equation in terms of the scale factor $a$, which may be solved to get the evolution of the universe filled with RCHDE. Moreover, using eqn.(\ref{denrc2}) and (\ref{flrweq2}) we get the pressure of the dark energy as,
\begin{equation}\label{pde}
p_{DE}=-\frac{1}{\kappa^{2}}\left[2\left\{\dot{H}+9\alpha\left(2H^{2}+\dot{H}\right)\right\}+\rho_{0}\kappa^{2}\left(1+z\right)^{3}+3\times6^{1/3}\lambda\left\{\tilde{\beta} H^{4}\left(2H^{2}+3\dot{H}\right)\right\}^{1/3}\right]
\end{equation}
The equation of state $w_{DE}=\frac{p_{DE}}{\rho_{DE}}$ may be obtained from eqns.(\ref{denrc2}) and (\ref{pde}).

%\begin{equation}\label{pde}
%p_{DE}=-\frac{1}{3\kappa^{2}a^{3}}\left[\kappa^{2}\rho_{0}+3a\left(6\alpha %\dot{a}^{2}+a\left(6\alpha-2\right)\ddot{a}-\frac{6^{1/3}\lambda}%{a}\left(\beta\dot{a}^{4}\left(\dot{a}^{2}-3a\ddot{a}\right)\right)^{1/3}\right)\right%]
%\end{equation}

Let us introduce the dimensionless density parameters as,
\begin{equation}\label{denpa}
\Omega_{m}\equiv
\frac{\kappa^{2}\rho_{m}}{3H^{2}},~~~~~~~~~~~\Omega_{DE}\equiv
\frac{\kappa^{2}\rho_{DE}}{3H^{2}}
\end{equation}
In terms of the above density parameters, the FLRW equation
(\ref{flrweq1}) becomes $\Omega_{m}+\Omega_{DE}=1$. Using the
energy density for dust we can write $\Omega_{m}$ as
$\Omega_{m}=\Omega_{m0}H_{0}^{2}/H^{2}a^{3}$, where $\Omega_{m0}$
is the present value of $\Omega_{m}$ given by
$\Omega_{m0}=\kappa^{2}\rho_{m0}/3H_{0}^{2}$. Similarly $H_{0}$ is
the present value of the Hubble function. Using this result we can
easily get the Hubble function as below,
\begin{equation}\label{hub}
H=\frac{H_{0}\sqrt{\Omega_{m0}}}{\sqrt{a^{3}\left(1-\Omega_{DE}\right)}}
\end{equation}
Here we will use $x=\ln a$ as the independent variable.
Differentiating eqn.(\ref{hub}) we have,
\begin{equation}\label{hdot}
\dot{H}=-\frac{H^{2}}{2\left(1-\Omega_{DE}\right)}\left[3\left(1-\Omega_{DE}\right)-\Omega_{DE}'\right]
\end{equation}
where prime represents derivative with respect to $x$. Using the
above equation in eqns.(\ref{ricfrw}) and (\ref{cubfrw}) we get,
\begin{equation}
R=-3H^{2}\left(1+\frac{\Omega_{DE}'}{1-\Omega_{DE}}\right)
\end{equation}
\begin{equation}
P=3\tilde{\beta}H^{6}\left(\frac{3\Omega_{DE}'}{1-\Omega_{DE}}-5\right)
\end{equation}
Using the above expressions for $R$ and $P$ in eqn.(\ref{denrc1})
we get,
\begin{equation}
\rho_{DE}=\frac{3H^{2}}{\kappa^2}\left[3^{1/3}\lambda
\left(\frac{\tilde{\beta}\left(5-5\Omega_{DE}-3\Omega_{DE}'\right)}{\Omega_{DE}-1}\right)^{1/3}+\frac{3\alpha\left(-1+\Omega_{DE}-\Omega_{DE}'\right)}{\Omega_{DE}-1}\right]
\end{equation}
Putting the above expression for energy density in (\ref{denpa})
we get the dimensionless energy density as,
\begin{equation}\label{energydiff}
\Omega_{DE}-3^{1/3}\lambda
\left(\frac{\tilde{\beta}\left(5-5\Omega_{DE}-3\Omega_{DE}'\right)}{\Omega_{DE}-1}\right)^{1/3}-\frac{3\alpha\left(-1+\Omega_{DE}-\Omega_{DE}'\right)}{\Omega_{DE}-1}=0
\end{equation}
The above differential equation governs the evolution of the
RCHDE for a flat universe with matter in the form of
dust. In the next section, we will use this equation to perform an observational data analysis on the RCHDE model to constrain the parameter space.

\section{Observational Data Analysis}
The equations derived in the previous section can be solved numerically for a range of redshift values to obtain $H(z)$ corresponding to various parameter combinations. These theoretical predictions will now be confronted with observational datasets through the chi-square minimization technique. This constitutes a statistical hypothesis test in which the null hypothesis asserts that the RCHDE model adequately fits the observational data. The best-fit parameters will be extracted by minimizing the chi-square function and will be used to produce the contour plots and $H(z)$ vs. $z$ evolution plots. To examine the observational validity of the Ricci-Cubic Holographic Dark Energy (RCHDE) model, we link the theoretical framework to actual cosmological observations by deriving predictions and comparing them to astrophysical data. Our main aim is to check if the model can reliably reproduce the expansion history of the universe, especially the variation of the Hubble parameter $H(z)$ with redshift.

The RCHDE framework is described through a set of interrelated equations involving the Hubble parameter $H(z)$ and the dark energy density parameter $\Omega_{\text{DE}}(z)$ given by eqns.(\ref{denpa}), (\ref{hub}) and (\ref{energydiff}). To determine how $\Omega_{\text{DE}}$ evolves with redshift, we solve numerically the nonlinear differential equation (\ref{energydiff}). The solution $\Omega_{\text{DE}}(z)$ is then plugged into equation (\ref{hub}) to reconstruct the theoretical Hubble parameter over the redshift range observed. To compare these theoretical outcomes with observational data, we utilize a chi-square ($\chi^2$) minimization method as part of a hypothesis testing framework. The null hypothesis posits that the RCHDE model, with appropriate parameters, can statistically explain the measured expansion rates. The chi-square is computed as
\begin{equation}
\chi^2 = \sum_{i=1}^N \frac{\left[ H_{\text{th}}(z_i) - H_{\text{obs}}(z_i) \right]^2}{\sigma_i^2},
\end{equation}
where $H_{\text{obs}}(z_i)$ denotes the observed Hubble parameter at redshift $z_i$, $\sigma_i$ its uncertainty, and $H_{\text{th}}(z_i)$ is the predicted value of the theoretical model. Minimizing $\chi^2$ yields the set of parameters $(\alpha, \lambda, \tilde{\beta})$ that best match the observations.

After identifying the best-fit parameters, the statistical confidence is evaluated using the likelihood function,
\begin{equation}
\mathcal{L} \propto \exp\left(-\frac{\chi^2}{2}\right),
\end{equation}
which is used to produce marginalized distributions and confidence contours (both one and two-dimensional) in the parameter space. These contours reveal the precision of parameter constraints and any correlations between them. In conclusion, by solving the fundamental dynamical equations of the model and rigorously comparing predictions with observational data through chi-square minimization, we can critically assess the compatibility of the RCHDE model with current cosmological measurements and place meaningful limits on its parameters.

\subsection{Results}
Now we proceed to present a thorough comparison of the RCHDE model with observational data from the universe. The analysis is based on two primary plot types:~ (i) \textbf{expansion history plots} that track how the universe’s expansion rate evolves over time, and ~ (ii) \textbf{statistical contour plots} that help identify the permissible ranges of the model’s parameters. A Markov-Chain Monte Carlo (MCMC) algorithm has been used for the analysis. The results are discussed below.

\subsubsection{Expansion History}

Expansion history plots illustrate the Hubble parameter $H(z)$, indicating how fast the universe expands at different redshifts $z$. Redshift serves as a proxy for looking back in time, with larger redshift values corresponding to earlier epochs in the evolution of the universe. For the RCHDE model, the theoretical predictions for $H(z)$ are compared against several observational datasets, including

\begin{itemize}

\item Direct Hubble parameter measurements and Cosmic Chronometers (which estimate expansion rates using the differential ages of passively evolving galaxies) (M. Moresco et al. 2016; C. Ranjit, P. Rudra \& S. Kundu 2021; P. Rudra \& Giri 2021),

\item Baryon Acoustic Oscillation (BAO) measurements, which act as cosmic distance indicators arising from matter density fluctuations in the early universe (W. J. Percival et al. 2010; F. Beutler et al. 2011; C. Blake et al. 2011; L. Anderson et al. 2012; N. G. Busca et al. 2013; L. Anderson et al. 2014; R. Tojeiro et al. 2014; M. Ata et al. 2018; T. M. C. Abbott et al. 2019; S. Alam et al. 2017; V. de Sainte Agathe et al. 2019; L. Kazantzidis et al. 2019; J. Hou et al. 2020; G. B. Zhao et al. 2020; D. Benisty \& D. Staicova 2020),

\item Dark Energy Spectroscopic Instrument (DESI) measurements, which provide precise constraints on the large-scale structure of the universe and the cosmic expansion history (DESI Collaboration et al. 2024; A. G. Adame et al. 2024a; A. G. Adame et al. 2024b).

\end{itemize}

Below we list the key observations from these comparisons:
\begin{itemize}
  \item At lower redshifts ($z \lesssim 1$), the behavior of the model closely matches that of the standard $\Lambda$CDM model, accurately describing the universe’s recent accelerated expansion.

  \item At higher redshifts ($z > 1$), the RCHDE model exhibits subtle yet meaningful departures from $\Lambda$CDM, owing to the influence of higher-order curvature modifications, which might point to new physical insights.
  
  \item Combining multiple datasets enhances the fit quality, with the model curve generally aligning well with measured data points and showing only minor deviations.
\end{itemize}

\subsubsection{Statistical Contours}

Statistical contour plots reveal constraints on the model’s main parameters: $\alpha$, $\lambda$, and $\tilde{\beta}$, $H_{0}$, and $\Omega_{m0}$  corresponding to a combination of Ricci scalar, cubic curvature scalar, and the non-linear interaction contributions, respectively.

The plots show confidence regions ($1\sigma$ and $2\sigma$) indicating likely parameter values:
\begin{itemize}
  \item Using single datasets tends to produce broader, more elongated confidence regions, reflecting greater uncertainty and possible correlations among parameters.
  \item When datasets are combined, the confidence regions become smaller and more symmetric, demonstrating tighter constraints and less parameter ambiguity.
\end{itemize}
\subsubsection{Best-Fit Parameters}

The model parameters are estimated using different observational datasets such as the Hubble parameter measurements, BAO data, DESI data and their combinations. The estimation procedure is performed by considering each dataset individually as well as their combined datasets in order to obtain stronger constraints on the model parameters.

The best-fit parameters are obtained through the chi-square minimization and Markov Chain Monte Carlo (MCMC) sampling technique within the Bayesian inference framework. The resulting best-fit values of the cosmological parameters are
\[
\alpha = 1.040789, \quad
\lambda = 0.620367, \quad
\tilde{\beta} = 0.975879, \quad
H_0 = 68.487369, \quad
\Omega_{m0} = 0.250573 .
\]

In Table~1 we present the estimated values of the parameters obtained from the different observational datasets and their combinations.

The interpretation of these values suggests the following:

\begin{itemize}
\item The value of $\alpha$ close to unity indicates that the Ricci scalar contribution plays a dominant role in the model dynamics, leading to behavior that closely resembles the $\Lambda$CDM model at late times.

\item The parameter $\lambda$ controls the contribution of the cubic curvature invariant. Its non–zero value indicates that higher-order curvature effects can influence the cosmic expansion history, particularly at earlier cosmological epochs.

\item The parameter $\tilde{\beta}$, which remains close to unity, suggests that the nonlinear curvature interactions remain well behaved and physically consistent within the model.
\end{itemize}

Overall, the combined observational and statistical analysis indicates that the RCHDE model provides a viable description of the expansion history of the universe. The model successfully reproduces the observed $H(z)$ behavior and remains consistent with current cosmological observations, while also allowing for additional curvature contributions beyond the standard $\Lambda$CDM framework.

\subsection{Description of likelihood Contours for parameters and $H(z)$ vs $z$ plots for different datasets}

Here we present and describe the likelihood contours generated for the model parameters using different observational datasets. In addition, we show the corresponding $H(z)$ versus $z$ plots obtained for these datasets. The likelihood contours provide probabilistic bounds on the estimated values of the free parameters of the model. The $H(z)$ vs $z$ plots illustrate the fit of the theoretical Hubble parameter with the observational data and allow a comparison with the predictions of the standard $\Lambda$CDM model. Each contour plot summarizes the posterior distributions of the five model parameters $\alpha$, $\lambda$, $\tilde{\beta}$, $H_{0}$, and $\Omega_{m0}$ obtained through Markov Chain Monte Carlo (MCMC) sampling. The analysis is performed using different observational datasets that constrain the model parameters within a Bayesian inference framework. The diagonal panels of the corner plots represent the marginalized posterior distributions for each parameter individually. The off-diagonal panels show the joint posterior distributions for each pair of parameters, where the contours correspond to the $68\%$ (1$\sigma$) and $95\%$ (2$\sigma$) credible regions. These contours illustrate the allowed parameter ranges as well as the correlations among the cosmological parameters.

\begin{figure}
    \centering
    \includegraphics[width=1\linewidth]{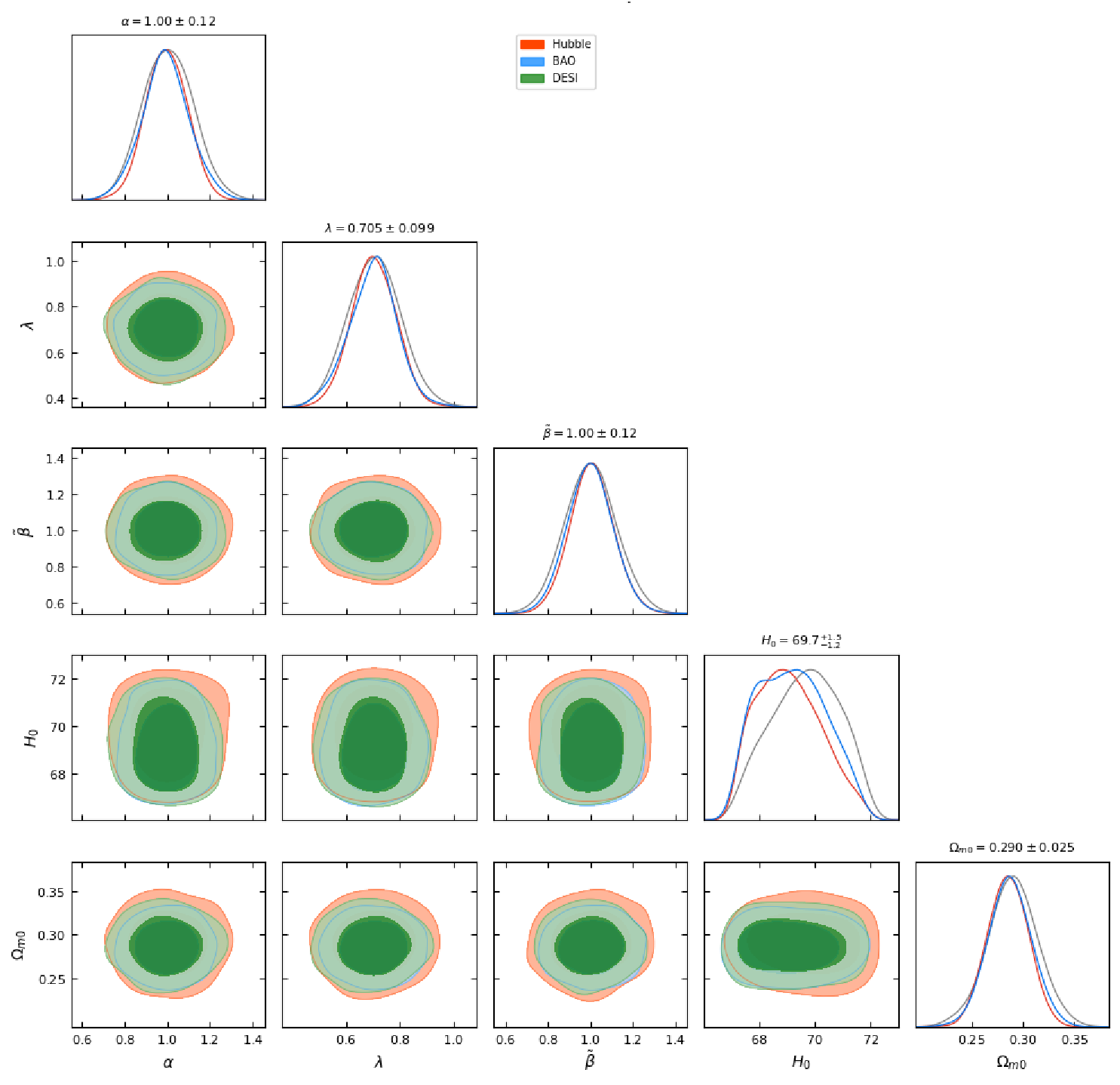}
    \caption{Joint and marginalized posterior distributions with combinations of datasets Hubble, BAO and DESI}
     \label{fig1}
\end{figure}

\begin{figure}
    \centering
    \includegraphics[width=1\linewidth]{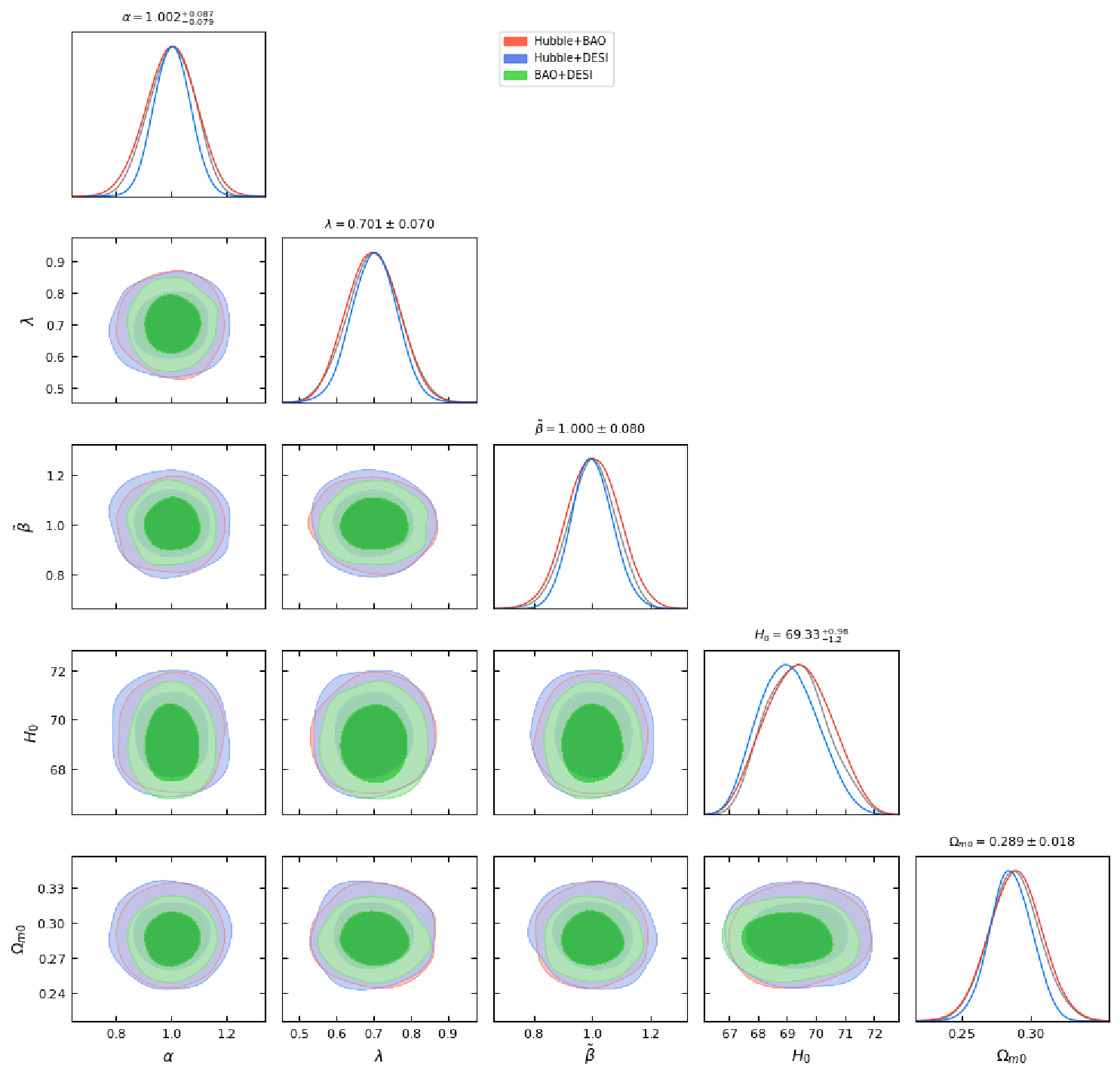}
    \caption{Joint and marginalized posterior distributions with dataset combination of Hubble+BAO, Hubble+DESI, BAO+DESI}
     \label{fig2}
\end{figure}

\begin{figure}
    \centering
    \includegraphics[width=1\linewidth]{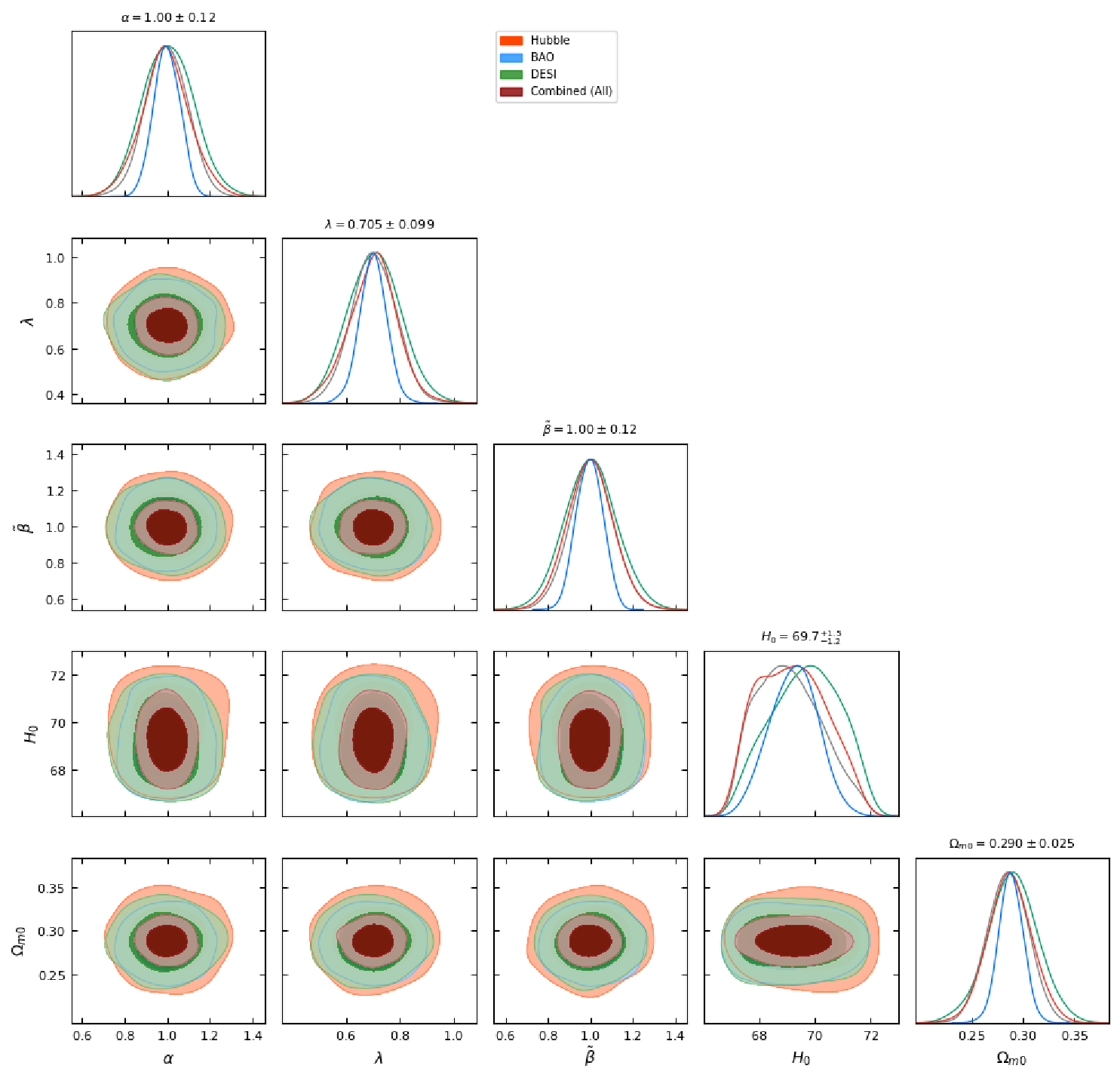}
    \caption{Joint and marginalized posterior distributions with combinations of datasets Hubble, BAO, DESI and Combined(Hubble+BAO+DESI)}
     \label{fig3}
\end{figure}

\begin{table}[H]
\centering
\caption{Parameter Constraints Summary }
\resizebox{\textwidth}{!}{%
\begin{tabular}{lccccc}
\hline\hline
Dataset & $\alpha$ & $\lambda$ & $\tilde{\beta}$ & $H_0$ [km s$^{-1}$ Mpc$^{-1}$] & $\Omega_{m0}$ \\
\hline
HUBBLE & 
$1.0049 \pm 0.1204$ & $0.7052 \pm 0.0988$ & $1.0022 \pm 0.1220$ & $69.67 \pm 1.25$ & $0.2900 \pm 0.0249$ \\

BAO & 
$0.9968 \pm 0.0980$ & $0.7032 \pm 0.0829$ & $1.0070 \pm 0.1032$ & $69.08 \pm 1.17$ & $0.2852 \pm 0.0199$ \\

DESI & 
$0.9937 \pm 0.1125$ & $0.6994 \pm 0.0902$ & $0.9993 \pm 0.1070$ & $69.18 \pm 1.22$ & $0.2872 \pm 0.0214$ \\

HUBBLE+BAO & 
$1.0017 \pm 0.0814$ & $0.7007 \pm 0.0700$ & $1.0004 \pm 0.0802$ & $69.33 \pm 1.04$ & $0.2885 \pm 0.0183$ \\

HUBBLE+DESI & 
$0.9998 \pm 0.0899$ & $0.6981 \pm 0.0694$ & $1.0040 \pm 0.0884$ & $69.40 \pm 1.11$ & $0.2893 \pm 0.0189$ \\

BAO+DESI & 
$1.0033 \pm 0.0656$ & $0.7018 \pm 0.0603$ & $1.0008 \pm 0.0693$ & $69.05 \pm 1.00$ & $0.2860 \pm 0.0149$ \\

HUBBLE+BAO+DESI & 
$0.9985 \pm 0.0612$ & $0.7001 \pm 0.0508$ & $0.9972 \pm 0.0626$ & $69.25 \pm 0.87$ & $0.2882 \pm 0.0119$ \\
\hline
\end{tabular}}
\label{tab:parameter_constraints}
\end{table}

\begin{figure}
    \centering
    \includegraphics[width=1\linewidth]{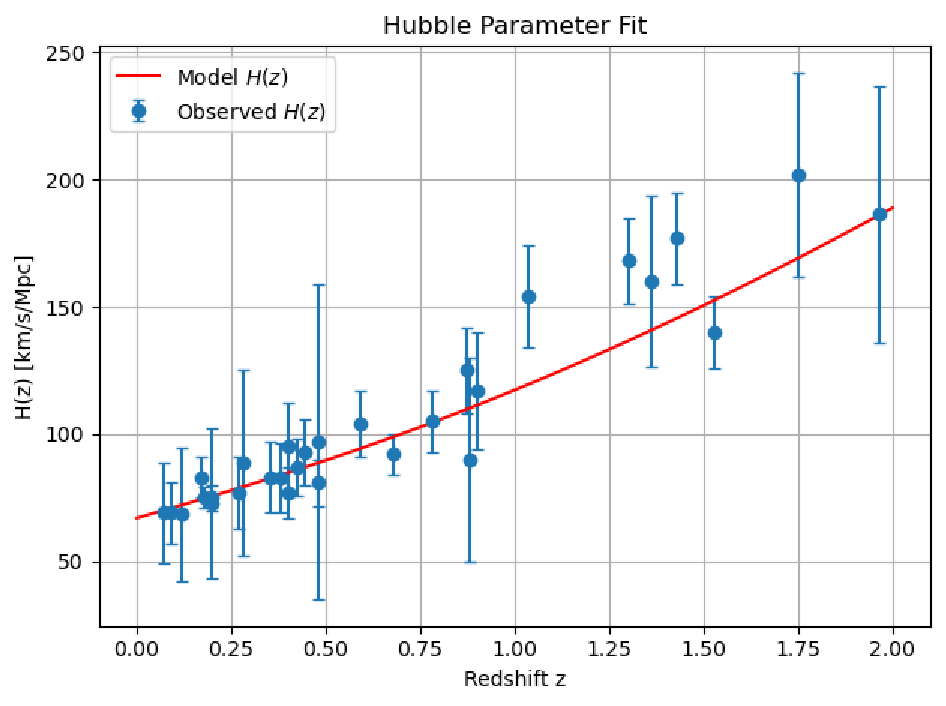}
    \caption{Plot shows the Hubble parameter fit (theoretical vs Hubble data) against the redshift parameter $z$}
    \label{fig6}
\end{figure}

\begin{figure}
    \centering
    \includegraphics[width=1\linewidth]{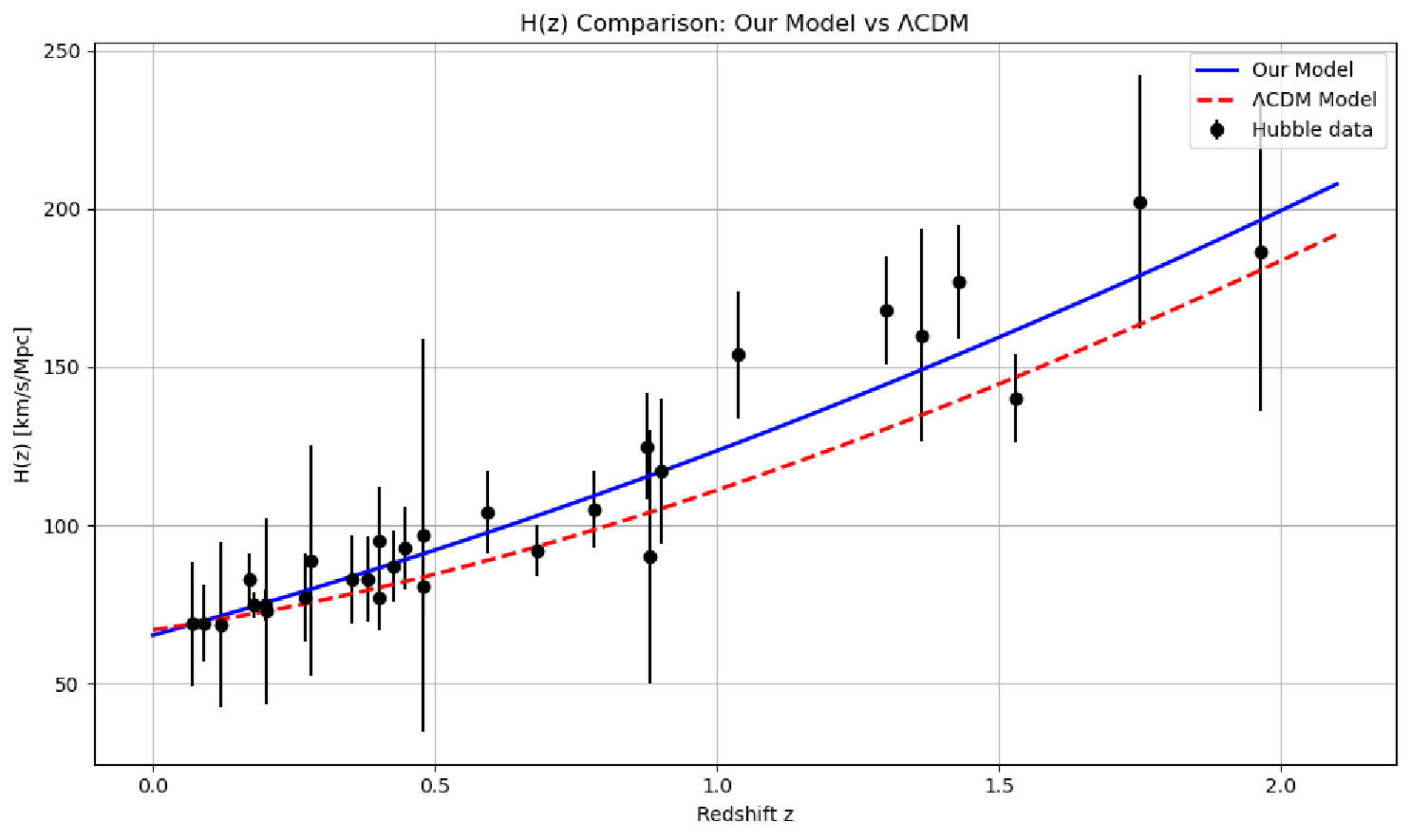}
    \caption{Plot showing the Hubble parameter fit (theoretical vs Hubble data) with a  comparison of the model with $\Lambda$CDM}
    \label{fig7}
\end{figure}

\begin{figure}
    \centering
    \includegraphics[width=1\linewidth]{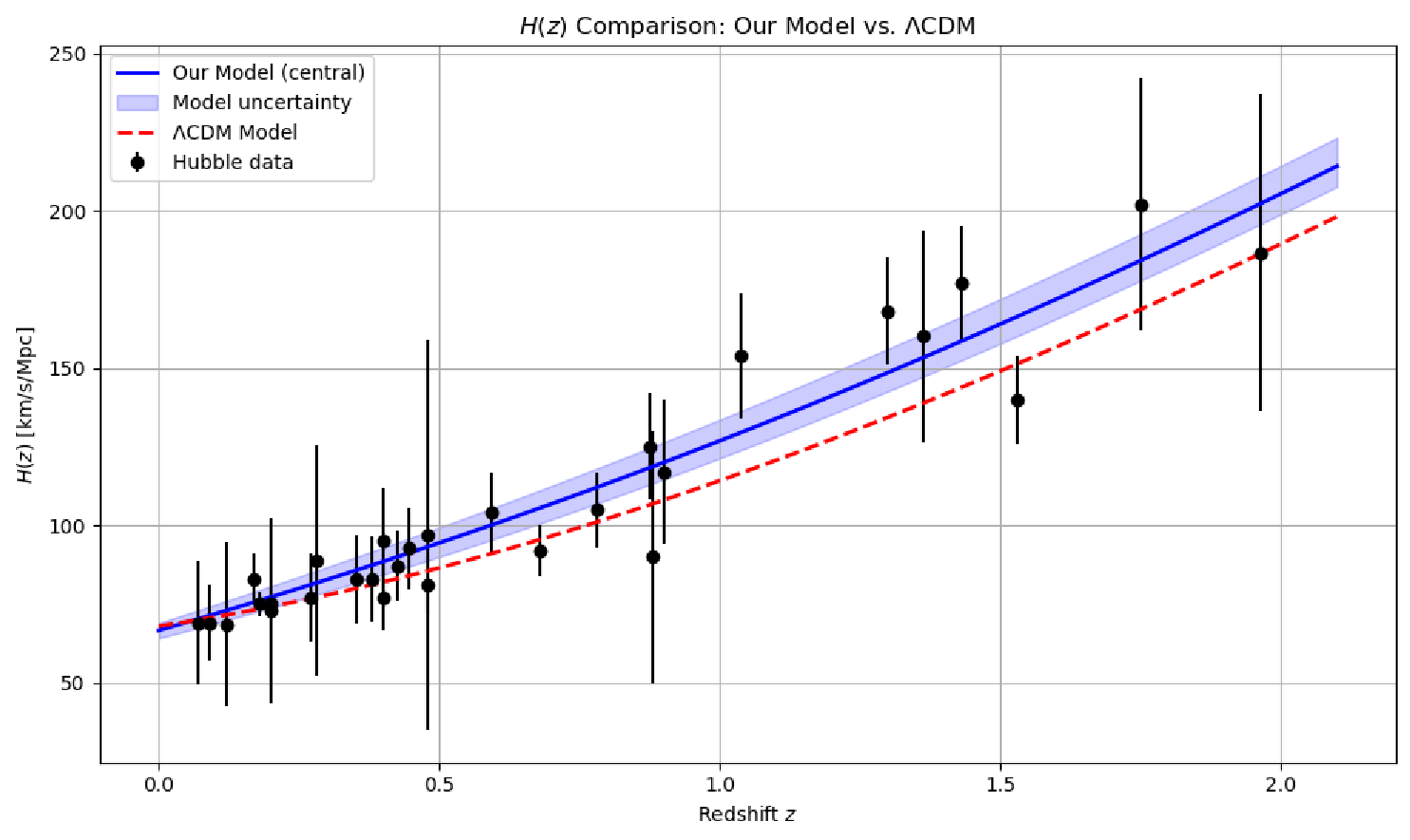}
    \caption{Plot showing the Hubble parameter fit (theoretical vs Hubble data) with a model $1 \sigma$ uncertainty region and a comparison of the model with $\Lambda$CDM}
    \label{fig8}
\end{figure}

\begin{figure}
    \centering
    \includegraphics[width=1\linewidth]{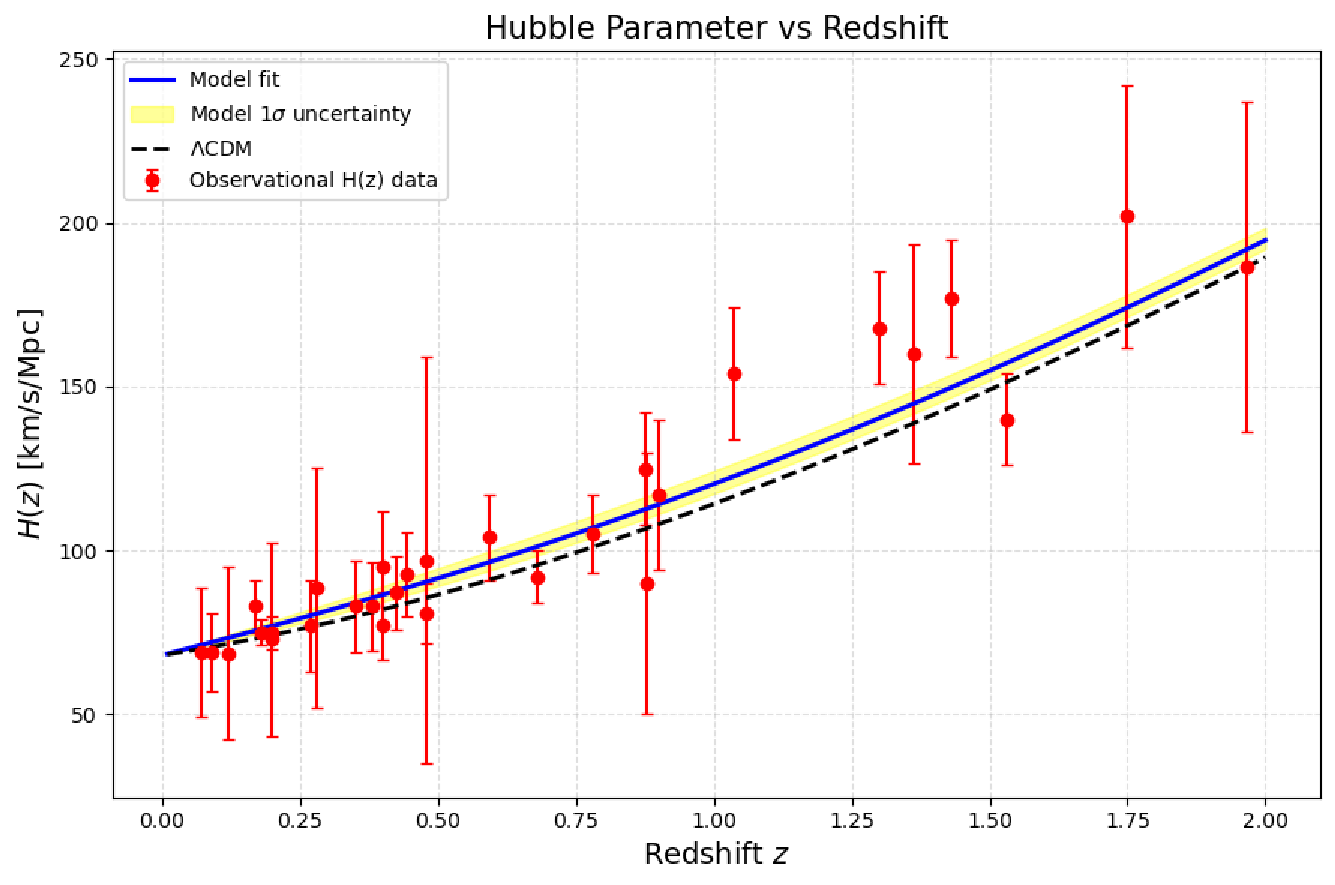}
    \caption{Plot showing the Hubble parameter fit (theoretical vs CC data) with a model uncertainty region and a comparison of the model with $\Lambda$CDM}
    \label{fig9}
\end{figure}

\begin{figure}
    \centering
    \includegraphics[width=1\linewidth]{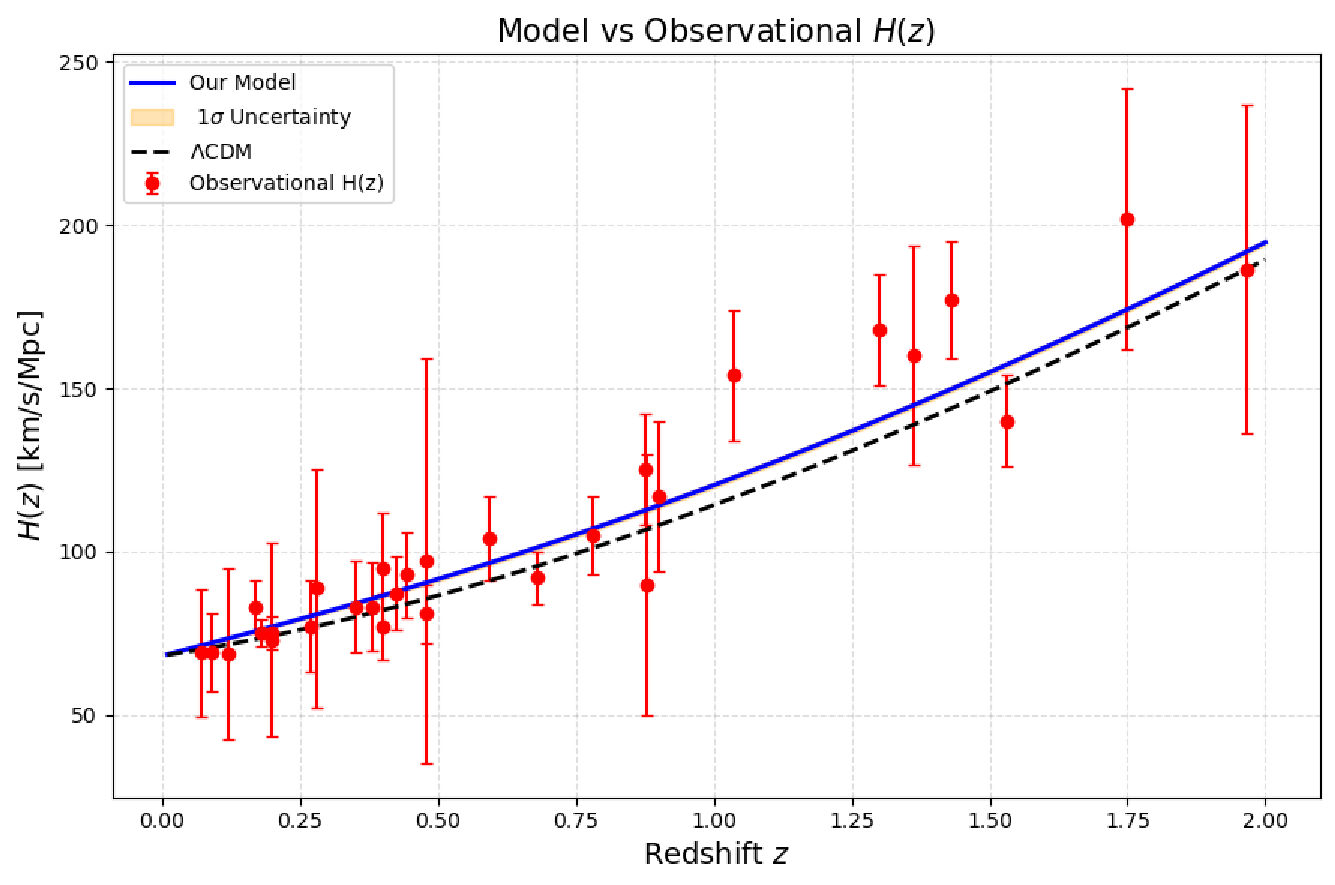}
    \caption{Plot showing the Hubble parameter fit (theoretical vs Hubble+CC data) with a model $1\sigma$ uncertainty region and a comparison of the model with $\Lambda$CDM}
    \label{fig11}
\end{figure}

Figure (\ref{fig1}) presents the joint and marginalized posterior distributions of the RCHDE model parameters obtained from the observational datasets consisting of Hubble parameter measurements, BAO, and DESI data. The diagonal panels represent the one-dimensional marginalized probability distributions for each parameter, while the off-diagonal panels display the corresponding two-dimensional confidence contours. The contours represent the $1\sigma$ and $2\sigma$ confidence regions, illustrating the allowed ranges of the parameters and their mutual correlations. The results show that the observational datasets provide meaningful constraints on the model parameters $\alpha$, $\lambda$, and $\tilde{\beta}$, along with the cosmological parameters $H_0$ and $\Omega_{m0}$. The distributions appear well localized, indicating that the RCHDE model parameters are consistent with the observational data within the expected confidence limits.

Figure (\ref{fig2}) shows the joint and marginalized posterior distributions of the RCHDE model parameters obtained using different combinations of observational datasets, namely Hubble+BAO, Hubble+DESI, and BAO+DESI. The diagonal panels illustrate the marginalized probability distributions of each parameter individually, whereas the off-diagonal panels display the two-dimensional joint posterior distributions with the corresponding $1\sigma$ and $2\sigma$ confidence contours. The comparison between different dataset combinations reveals how the inclusion of additional observational data improves the constraints on the cosmological parameters. In particular, combining datasets leads to narrower confidence regions, indicating a reduction in parameter degeneracies and a more precise determination of the model parameters.

Figure (\ref{fig3}) illustrates the joint and marginalized posterior distributions of the RCHDE model parameters obtained using the individual datasets Hubble, BAO, and DESI, along with their combined dataset (Hubble+BAO+DESI). The diagonal panels show the marginalized distributions for each parameter, while the off-diagonal panels present the corresponding two-dimensional confidence contours representing the $1\sigma$ and $2\sigma$ credible regions. The combined dataset provides tighter and more symmetric confidence regions compared to the individual datasets, indicating stronger constraints on the cosmological parameters. This demonstrates that the simultaneous use of multiple observational datasets significantly improves the robustness and reliability of the parameter estimation for the RCHDE model.

Figure (\ref{fig6}) presents the comparison between the observed Hubble parameter $H(z)$ data and the predictions from our theoretical cosmological model. The dataset consists of empirical $H(z)$ measurements as a function of redshift $z$, compiled from recent observational studies. In the plot, observed $H(z)$ values with $1\sigma$ error bars are depicted by blue filled circles, while the solid red line shows the prediction from the best-fit model, obtained via least-squares fitting or Markov Chain Monte Carlo (MCMC) parameter estimation. The tight agreement between the model curve and the observed data across the sampled redshift range demonstrates the model's suitability for capturing the Hubble expansion history. The observational data display the expected trend of increasing $H(z)$ with redshift, and the model provides a smooth analytical fit closely tracking the mean values and uncertainties of the data points. This comparison is central to validating the cosmological model and assessing its ability to reproduce the universe’s expansion dynamics. 

Figure (\ref{fig7}) presents a comparative analysis between the predictions of our cosmological model and the standard $\Lambda$CDM model against observational Hubble parameter $H(z)$ measurements. The data points, along with their uncertainties, are derived from direct Hubble parameter observations. In the plot, the solid blue line corresponds to the prediction from our cosmological model, while the red dashed line denotes the canonical $\Lambda$CDM prediction. Observational $H(z)$ data and their $1\sigma$ error bars are shown as black circles. This visualization allows for a straightforward performance comparison between the two theoretical models and the empirical data across the redshift range $0 < z < 2.1$. Our model demonstrates a strong fit to the Hubble parameter measurements, closely following the observed trend at both low and moderate redshifts and successfully capturing the increase of $H(z)$ with redshift. The $\Lambda$CDM model also generally reproduces the data but shows modest deviations at higher redshift. These results highlight the utility of the empirical Hubble parameter data for discriminating between cosmological scenarios. From this, we see that our model is a better fit to the Hubble data compared to the $\Lambda$CDM model. 

Figure (\ref{fig8}) displays a direct comparison between the predictions from our cosmological model and the standard $\Lambda$CDM model against observational Hubble parameter $H(z)$ data, just like figure (\ref{fig7}). Here we have shown a range (light blue shaded area around the blue central line) depicting the uncertainty of the model. Our model demonstrates excellent agreement with the empirical data across the full redshift range ($0 < z < 2.1$), with most datapoints falling within the $1\sigma$ uncertainty band. One important thing to note is that the $\Lambda$CDM line lies outside this uncertainty band at moderate/higher redshifts, but as we approach the lower redshifts (late universe), the line slowly gets into the uncertainty band and finally almost coincides with our model.

In Figure (\ref{fig9}), we compare the predictions of our cosmological model to observational Hubble parameter data obtained from cosmic chronometers (CC). The analysis employs state-of-the-art Hubble parameter measurements from recent cosmic chronometer samples. The solid blue line presents the best-fit prediction from our model, with the yellow band illustrating the associated $1\sigma$ confidence interval as determined from Markov Chain Monte Carlo (MCMC) posterior sampling. The standard $\Lambda$CDM prediction is shown by the black dashed line. Observational $H(z)$ values derived from CC data are displayed as red circles with error bars, highlighting measurement uncertainties across the sampled redshift range. Our model provides a robust fit to the CC-derived $H(z)$ data over all probed redshifts, with most observational points lying within the $1\sigma$ uncertainty region. While the $\Lambda$CDM model broadly replicates the observed trend, small differences are visible especially at higher redshifts, underscoring the discriminating power of precise CC measurements.

Figure (\ref{fig11}) illustrates the comparative analysis between the predictions of our cosmological model and the standard $\Lambda$CDM scenario against observations from Hubble parameter $H(z)$ data, combining results from both Hubble measurements and cosmic chronometers (CC). The dataset includes recent $H(z)$ values and uncertainties compiled from contemporary CC and Hubble samples. Our model provides an excellent fit to the combined Hubble and CC dataset across $0 < z < 2$, with the majority of data points lying within the model’s $1\sigma$ uncertainty envelope. While the standard $\Lambda$CDM prediction captures the global trend, small deviations become visible at higher redshifts, demonstrating the value of joint datasets for testing model robustness and distinguishing subtle cosmological effects.

\section{Hubble Tension Analysis of our Model}
The Hubble tension is one of the biggest unresolved problems in modern cosmology. It refers to a persistent disagreement in the value of the Hubble constant $H_0$ (the current expansion rate of the universe) when measured using different methods, interpreted within the framework of the $\Lambda$CDM model. This disagreement arises from two conflicting measurements- the early universe CMB-based measurements using Planck data, and the late universe local measurements from Cepheid variables and Type Ia Supernova. From the early-universe measurements we get $H_0=67$ km/s/Mpc, and from the late-universe measurements we get $H_0=73$ km/s/Mpc. (Planck collaboration 2018; L. Verde et al. 2019; A. G. Riess et al. 2022). The difference between these measurements exceeds the level expected from statistical uncertainties and currently corresponds to a discrepancy of approximately $4$--$5\sigma$. This tension may indicate the presence of unknown systematic errors or point toward new physics beyond the standard $\Lambda$CDM cosmological model (Planck collaboration 2018; L. Verde et al. 2019; A. G. Riess et al. 2022).

To quantify the level of disagreement between two independent measurements of the Hubble constant, we adopt the standard Gaussian estimator (L. Verde et al. 2019):
\begin{equation}
T = \frac{\left| H_0^{(1)} - H_0^{(2)} \right|}{\sqrt{\sigma_1^2 + \sigma_2^2}},
\end{equation}
where $H_0^{(1)}$ and $H_0^{(2)}$ represent the two measurements being compared, and $\sigma_1$, $\sigma_2$ are their corresponding uncertainties. The quantity $T$ measures the tension in units of standard deviation ($\sigma$).

In this work, we assess the consistency of our model with respect to the standard $\Lambda$CDM cosmology by computing the dataset-wise Hubble tension. For each dataset, the $\Lambda$CDM value is taken as the reference and is compared with the corresponding value obtained from our model.

\begin{table}[H]
\centering
\caption{Dataset-wise Hubble tension between the proposed RCHDE model and the $\Lambda$CDM reference. All tensions are found to be below $0.05\sigma$, indicating excellent agreement.}
\resizebox{\textwidth}{!}{%
\begin{tabular}{lccccc}
\hline\hline
Dataset & $H_0^{\Lambda \mathrm{CDM}}$ & $H_0^{\mathrm{model}}$ & $\Delta H_0$ & $\sigma_{\mathrm{tot}}$ & Tension ($\sigma$) \\
\hline
HUBBLE & $69.70 \pm 1.28$ & $69.67 \pm 1.25$ & 0.03 & 1.79 & $0.02\sigma$ \\
BAO & $69.02 \pm 1.18$ & $69.08 \pm 1.17$ & 0.06 & 1.66 & $0.04\sigma$ \\
DESI & $69.16 \pm 1.20$ & $69.18 \pm 1.22$ & 0.02 & 1.71 & $0.01\sigma$ \\
HUBBLE+BAO & $69.32 \pm 1.04$ & $69.33 \pm 1.04$ & 0.01 & 1.47 & $0.01\sigma$ \\
HUBBLE+DESI & $69.46 \pm 1.14$ & $69.40 \pm 1.11$ & 0.06 & 1.59 & $0.04\sigma$ \\
BAO+DESI & $69.01 \pm 0.97$ & $69.05 \pm 1.00$ & 0.04 & 1.40 & $0.03\sigma$ \\
HUBBLE+BAO+DESI & $69.21 \pm 0.87$ & $69.25 \pm 0.87$ & 0.04 & 1.23 & $0.03\sigma$ \\
\hline
\end{tabular}%
}
\label{tab:hubble_tension_lcdm}
\end{table}

\begin{figure}
    \centering
    \includegraphics[width=1\linewidth]{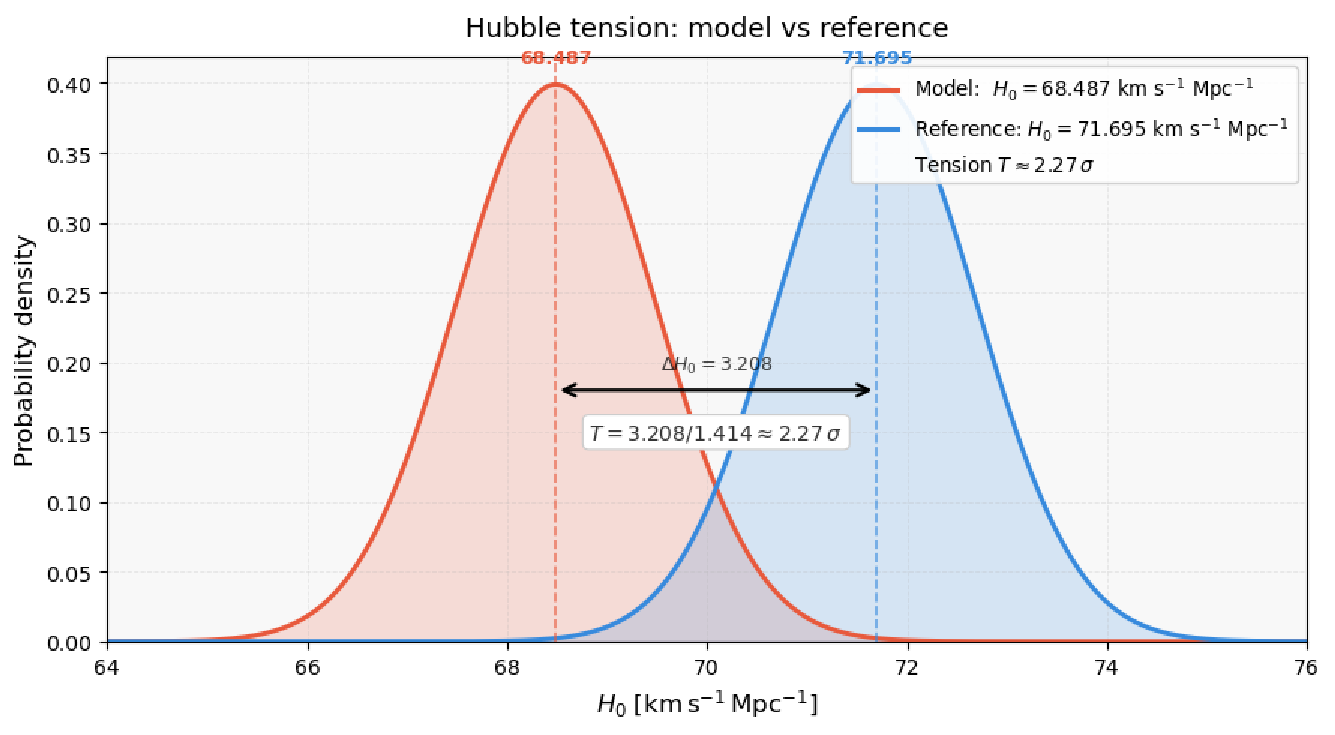}
    \caption{Illustration of the Hubble tension between the best-fit value obtained from the RCHDE model and the reference value from $\Lambda$CDM}
    \label{fig:placeholder}
\end{figure}

\subsection{Hubble Tension from Best-Fit Values}
For the present analysis, we consider the reference value coming from the $\Lambda$ CDM model for the datasets used in this work. This value is given by 
\begin{equation}
H_0^{\mathrm{ref}} = 71.695493,
\end{equation}
The best-fit value obtained from our RCHDE model is
\begin{equation}
H_0^{\mathrm{model}} = 68.487369.
\end{equation}
The resulting Hubble tension is given by
\begin{equation}
T \approx \frac{3.208124}{1.414} \approx 2.27\sigma.
\end{equation}

\noindent
Therefore, the best-fit value obtained in our model exhibits a moderate tension of approximately $2.3\sigma$ with respect to the reference value. While this indicates a noticeable discrepancy, it remains significantly lower than the $\sim 5\sigma$ tension typically reported between early- and late-Universe measurements, suggesting a partial alleviation of the Hubble tension.
 % end color blue

%%%%%%%%%%%%%%%%%%%%%%%%%%%%%%%%%%%%%%%%%%%%%%%%%%%%%%%%%%%%%%%%%%
\section{Enhanced Machine Learning Analysis of the RCHDE Model}
%%%%%%%%%%%%%%%%%%%%%%%%%%%%%%%%%%%%%%%%%%%%%%%%%%%%%%%%%%%%%%%%%%
To complement the theoretical modeling of the RCHDE model using the MCMC posterior sampling method, we carried out an enhanced machine learning (ML) analysis using observational Hubble parameter data. This approach serves to validate the model's predictive power through independent, data-driven regression techniques. Machine learning is used in this context because modern cosmological datasets are both large and complex, containing high-dimensional, nonlinear relationships between observational quantities and cosmological parameters. ML techniques can efficiently uncover these complex correlations and provide accurate mappings from data to theoretical models, which can be challenging or computationally expensive for traditional statistical approaches. The major advantages of ML in cosmological analysis are its capacity to reduce computational time for parameter estimation, increase the precision of constraints, and robustly incorporate information from diverse data sources. Algorithms such as artificial neural networks, support vector regression, and ensemble methods are able to quickly explore large parameter spaces, leading to more efficient and potentially more accurate cosmological inferences. Furthermore, ML can learn from simulated or theoretical universes, detect subtle signatures of new physics, and facilitate a synergy between data-driven and theory-driven approaches.

\subsection{Machine Learning Implementation and Evaluation}

We employed six supervised ML regression algorithms to reconstruct the Hubble parameter function and compare their performance with the theoretical RCHDE model:
\begin{enumerate}
    \item Enhanced Linear Regression,
    \item Physics-Informed Linear Regression,
    \item Enhanced Artificial Neural Network (ANN),
    \item Enhanced Support Vector Regression (SVR),
    \item Enhanced Random Forest Regression,
    \item Gradient Boosting Regression.
\end{enumerate}

These models were trained using the 20-point training set and evaluated on the 10-point test set. Performance was assessed using the coefficient of determination (\( R^2 \)), root mean square error (RMSE), chi-squared statistics, reduced chi-squared, and mean absolute deviation in the parameter \( \alpha \).

\subsubsection*{Performance Summary}

Table~\ref{tab:ml_summary} presents a comparative summary of all models. The Enhanced SVR algorithm outperformed the others in terms of test \( R^2 \), chi-squared, and generalization ability.

\begin{table}[H]
\centering
\caption{Performance Comparison of Machine Learning Algorithms}
\label{tab:ml_summary}
\begin{tabular}{lcccccc}
\toprule
\textbf{Algorithm} & \textbf{Train $R^2$} & \textbf{Test $R^2$} & \textbf{Test RMSE} & \textbf{$\chi^2$} & \textbf{$\chi^2_\nu$} & \textbf{Mean $|\Delta\alpha|$} \\
\midrule
Enhanced SVR            & 0.9280 & 0.8690 & 15.80 & 14.27 & 0.4921 & 0.0725 \\
Enhanced ANN            & 0.9560 & 0.8451 & 17.18 & 15.85 & 0.5465 & 0.0702 \\
Enhanced Linear         & 0.9044 & 0.8266 & 18.17 & 18.48 & 0.6373 & 0.0913 \\
Physics-Informed Linear & 0.8959 & 0.8154 & 18.75 & 19.64 & 0.6771 & 0.0847 \\
Enhanced Random Forest  & 0.9806 & 0.7861 & 20.19 & 17.58 & 0.6062 & 0.0677 \\
Gradient Boosting       & 0.9654 & 0.7772 & 20.60 & 19.92 & 0.6868 & 0.0816 \\
\bottomrule
\end{tabular}
\end{table}

\subsubsection*{Best Performing Model: Enhanced SVR}

The Enhanced SVR model exhibited the best balance between bias and variance:
\begin{itemize}
    \item \( R^2_{\text{test}} = 0.8690 \),
    \item RMSE = 15.80 km/s/Mpc,
    \item \( \chi^2 = 14.27 \), closely matching the theoretical model,
    \item Reduced \( \chi^2_\nu = 0.4921 \),
    \item Mean absolute deviation in \( \alpha \) = 0.0725.
\end{itemize}

These metrics suggest that the SVR model not only captures the underlying cosmological behavior but also aligns closely with the theoretical expectations of the RCHDE scenario.
 \begin{figure}
     \centering
     \includegraphics[width=1\linewidth]{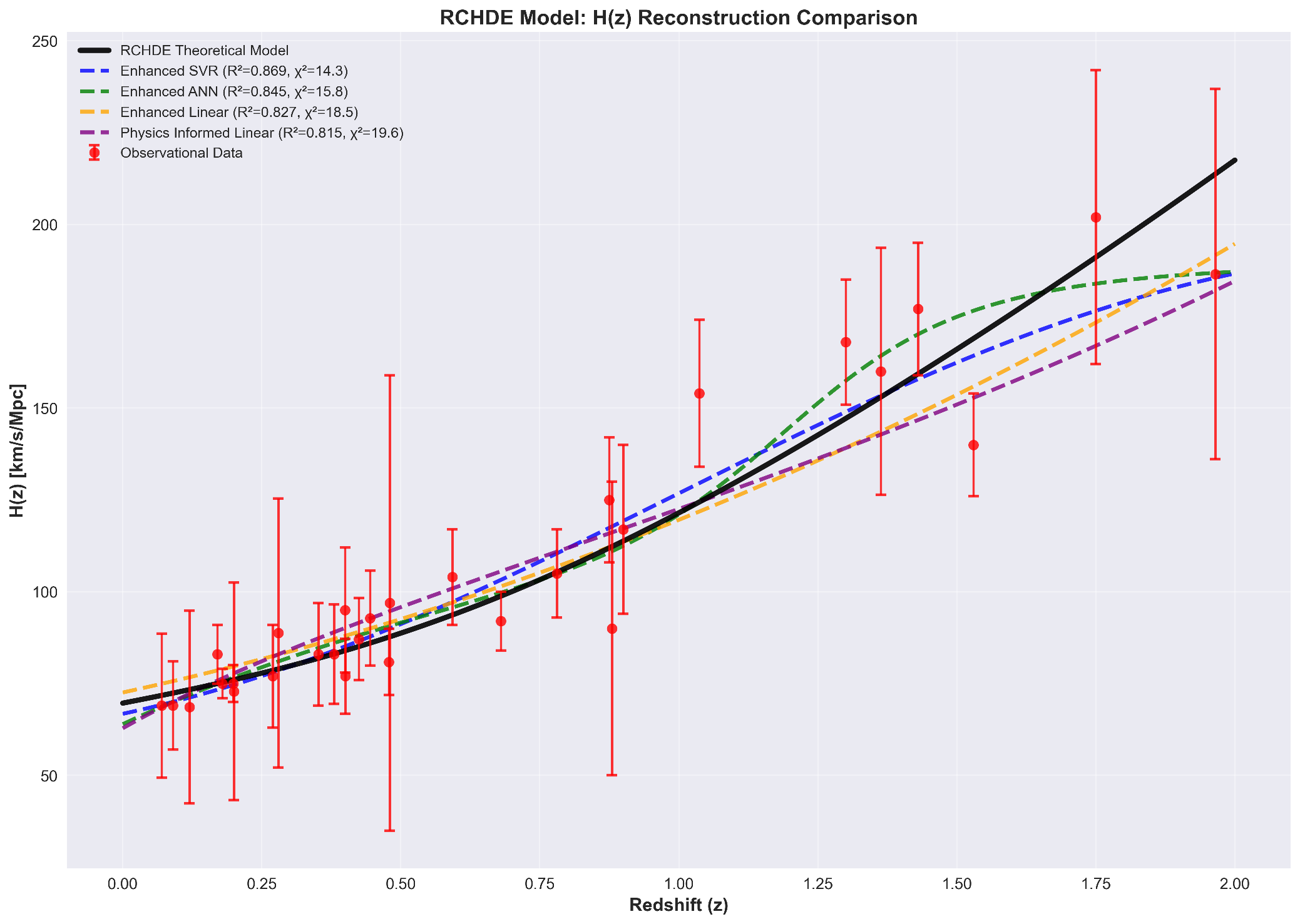}
     \caption{Machine learning regression fit to observational Hubble parameter $H(z)$ data using the best-fit Ricci-Cubic Holographic Dark Energy (RCHDE) model. The solid black line shows the RCHDE theoretical prediction, while data points (red circles) correspond to measured $H(z)$ values with $1\sigma$ error bars. The regression fits from ML algorithms are shown using dashed lines. The close alignment illustrates the model's accuracy in capturing the cosmic expansion history.}
     \label{fig12}
 \end{figure}
 \begin{figure}
     \centering
     \includegraphics[width=1\linewidth]{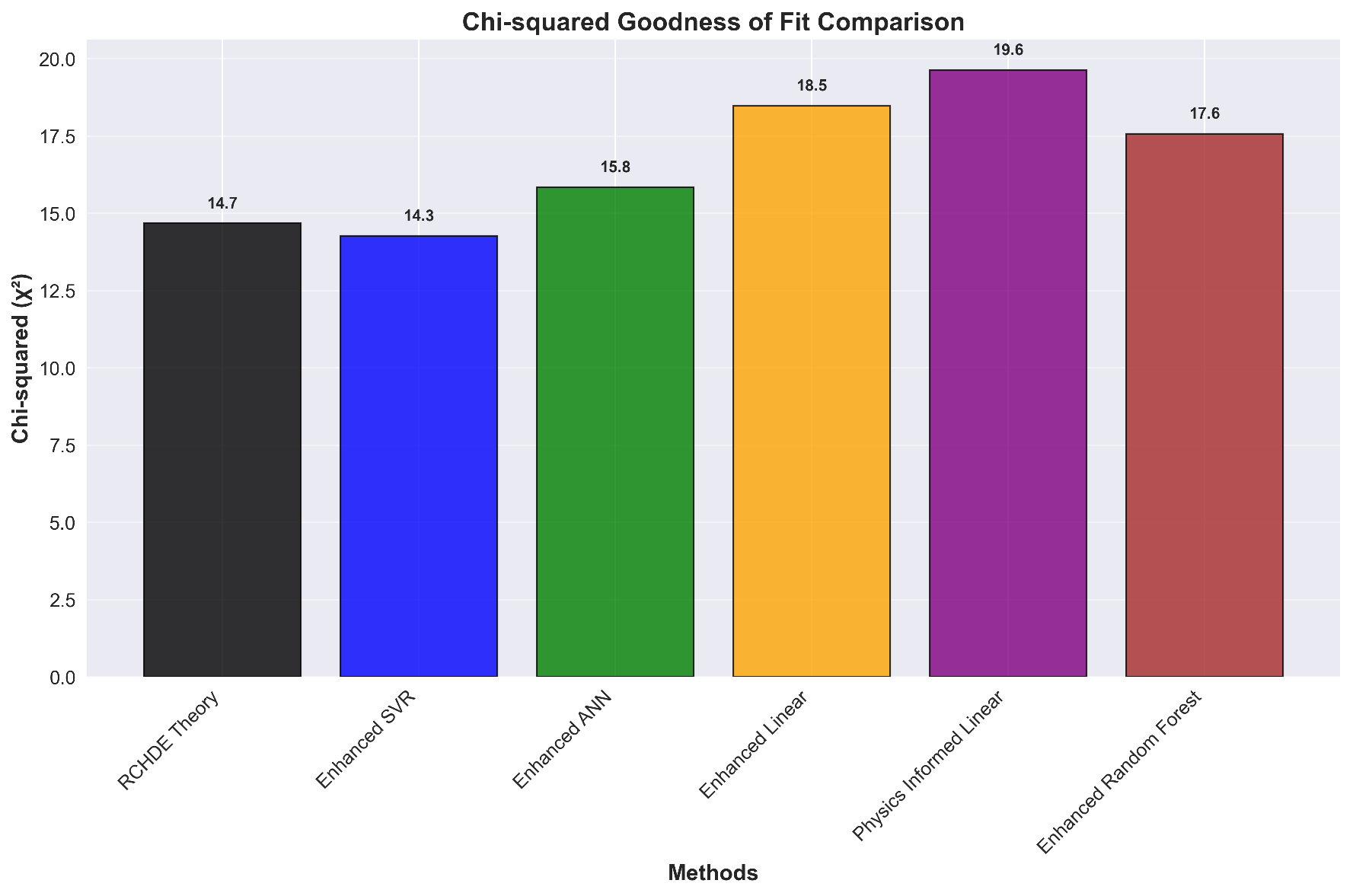}
     \caption{$\chi^2$ goodness of fit for different ML techniques in a bar diagram showing their comparative values.}
     \label{fig13}
 \end{figure}
 \begin{figure}
     \centering
     \includegraphics[width=1\linewidth]{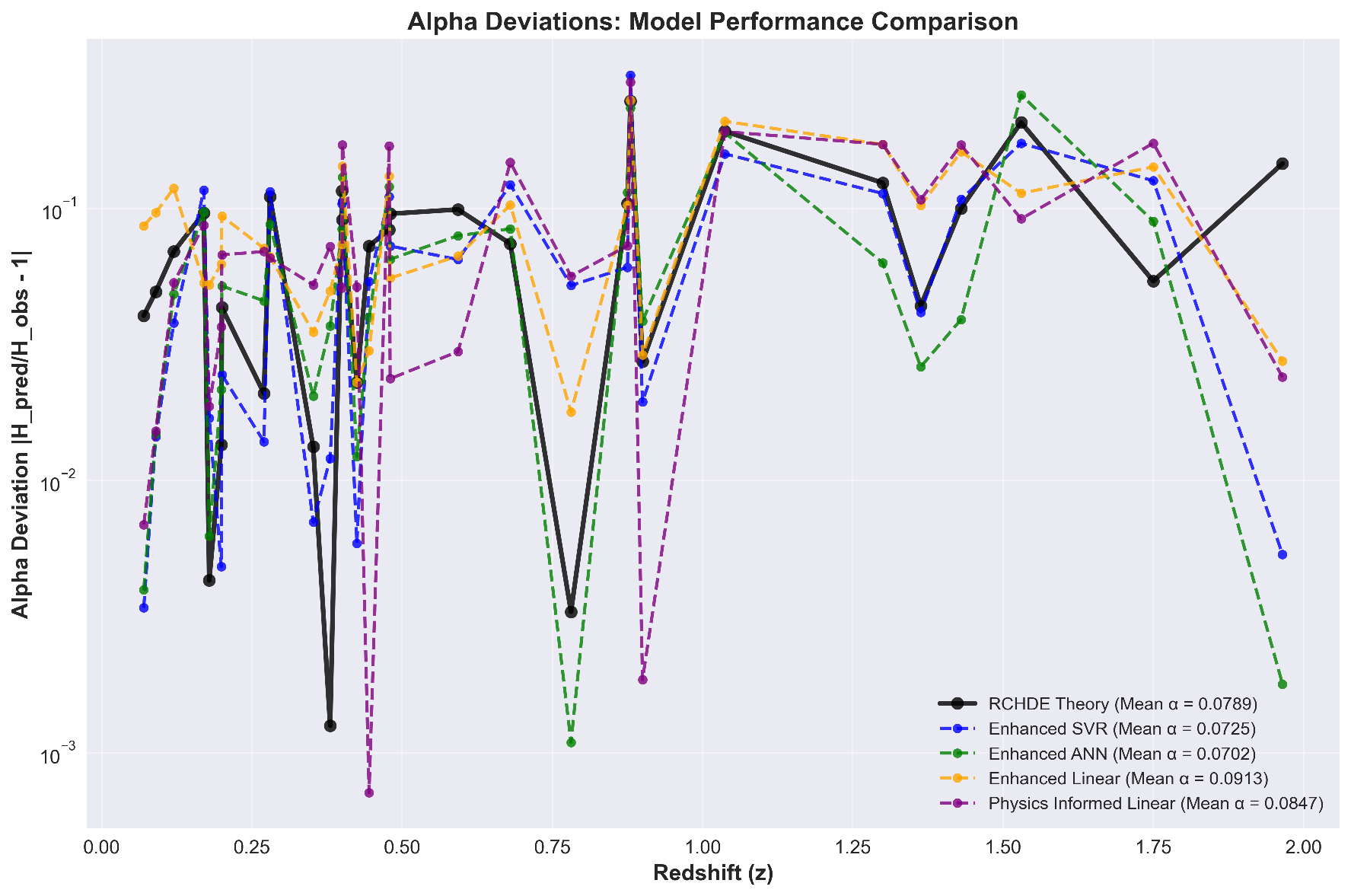}
     \caption{Comparison of the absolute relative deviation $|H_{\mathrm{pred}}/H_{\mathrm{obs}} - 1|$ of the Hubble parameter from different predictive models as a function of redshift. Curves represent the RCHDE theory, Enhanced Support Vector Regression (SVR), Enhanced Artificial Neural Network (ANN), Enhanced Linear Regression, and Physics-Informed Linear Regression. The plot demonstrates that machine learning models, especially Enhanced SVR and ANN, exhibit deviations comparable to the RCHDE theoretical model, indicating strong predictive performance.}
     \label{fig14}
 \end{figure}
\begin{figure}
    \centering
    \includegraphics[width=1\linewidth]{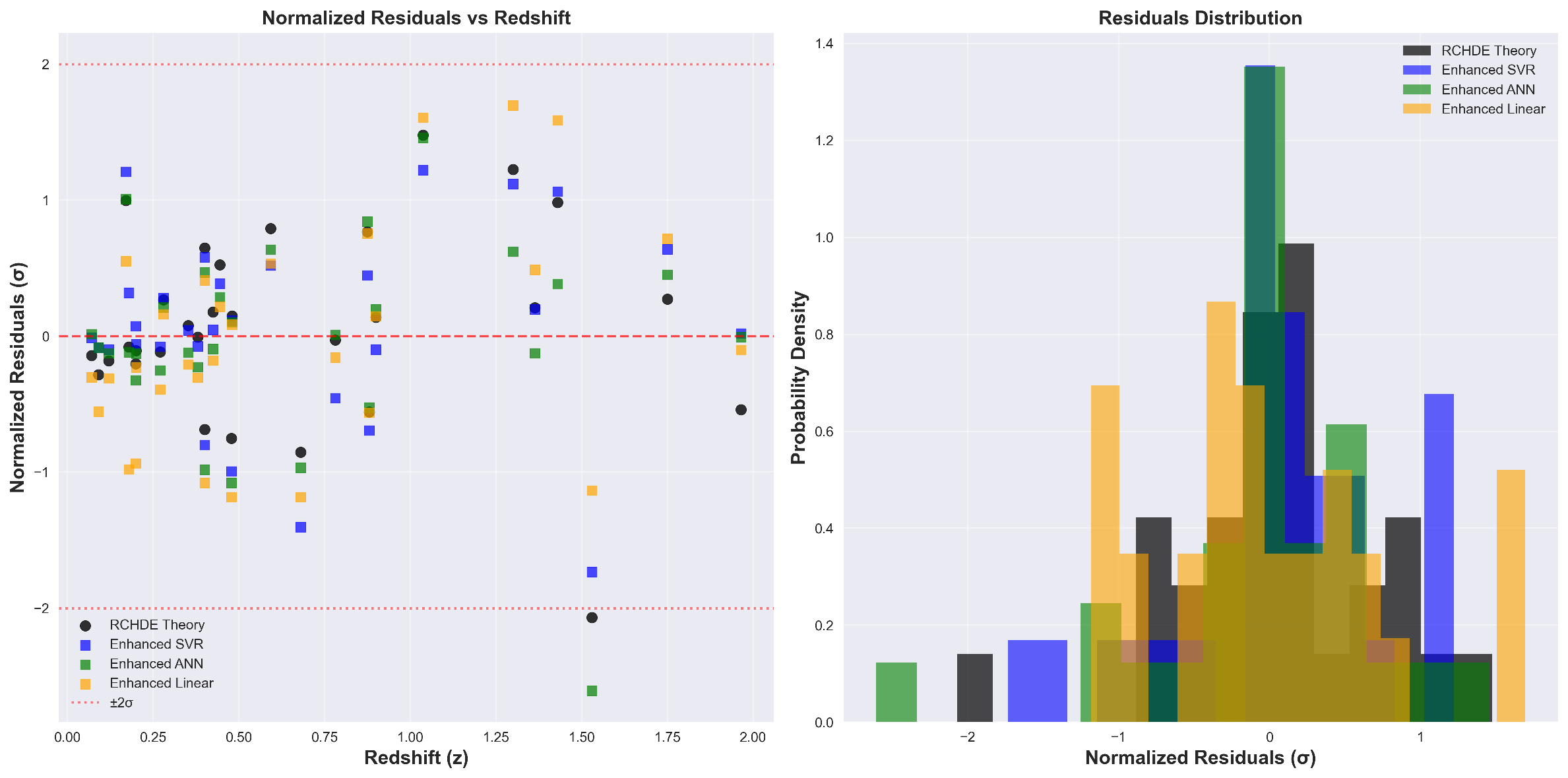}
    \caption{Distribution of normalized residuals (in units of $\sigma$) between predicted and observed $H(z)$ for different models: RCHDE Theory, Enhanced SVR, Enhanced ANN, and Enhanced Linear Regression. The histogram illustrates that all models yield residuals centered around zero with most data points within $\pm1\sigma$, indicating unbiased predictions and robust model fits to the observational data.}
    \label{fig15}
\end{figure}

\begin{figure}
    \centering
    \includegraphics[width=1\linewidth]{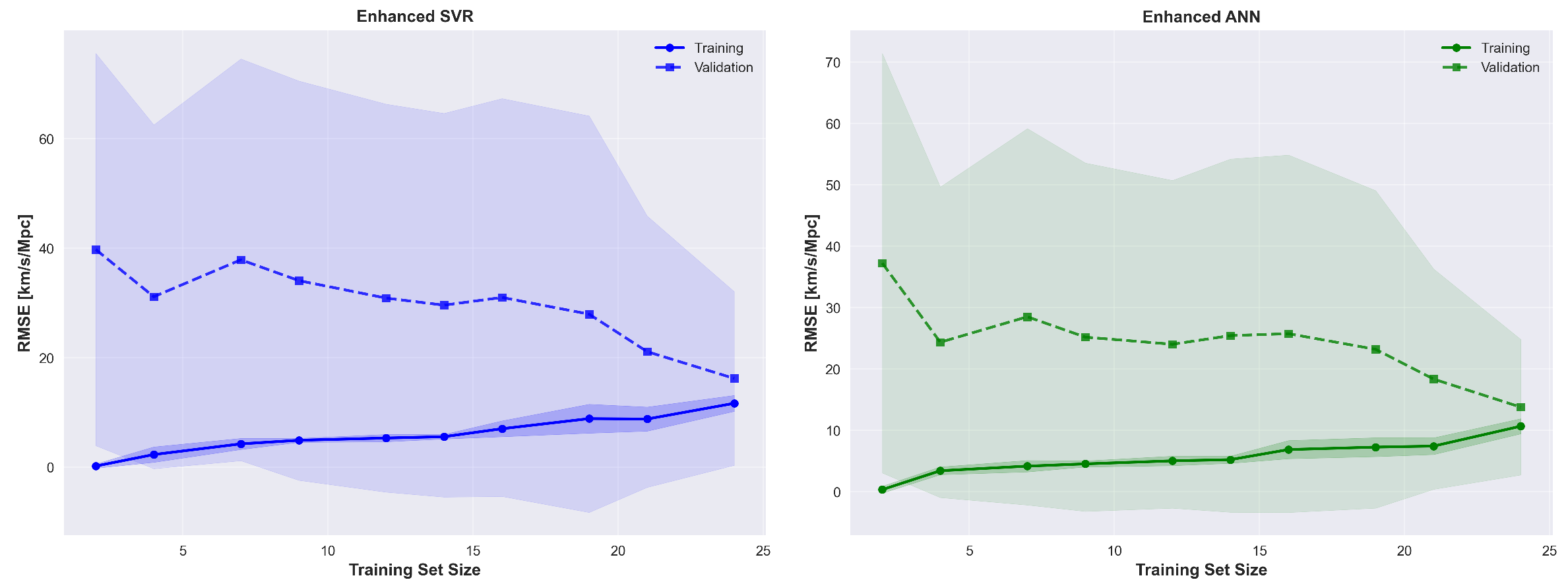}
    \caption{Probability density distribution of normalized residuals, comparing the RCHDE theoretical model with enhanced machine learning regressors (SVR, ANN, and Linear Regression). The distributions show that the majority of residuals for all models are tightly clustered around zero, confirming consistency and high accuracy in reconstructing the Hubble expansion history from observational data.}
    \label{fig16}
\end{figure}
\begin{figure}
    \centering
    \includegraphics[width=1\linewidth]{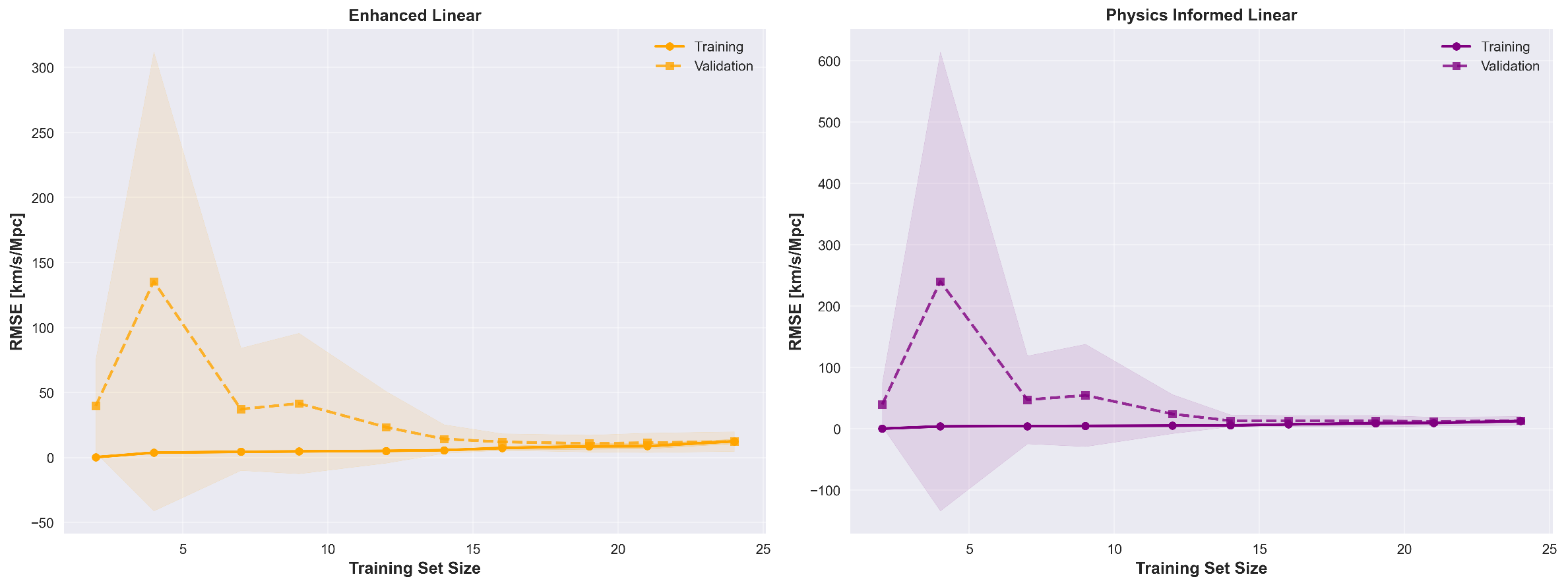}
    \caption{Summary comparison of mean absolute deviations in the model parameter $\alpha$ among the tested regression approaches and the RCHDE theory.}
    \label{fig17}
\end{figure}

In this work, we perform an enhanced machine learning (ML) analysis to validate and complement the theoretical modeling of the Ricci-Cubic Holographic Dark Energy (RCHDE) model using observational Hubble parameter data. The dataset employed comprises 30 observed values of the Hubble parameter \(H(z)\), covering the redshift range \(z = 0.070\) to \(z = 1.965\), with \(H(z)\) values ranging from 68.6 to 202.0 km/s/Mpc. This dataset is divided into a training set containing 20 data points and a test set of 10 data points. The theoretical RCHDE model parameters were optimized via chi-square minimization, yielding best-fit values of the Hubble constant \(H_0 = 69.68\) km/s/Mpc, matter density parameter \(\Omega_{m0} = 0.360\), interaction parameter \(\alpha = 0.765\), normalized coupling \(\tilde{\beta} = 1.000\), and evolution parameter \(\lambda = 0.700\), with a fit corresponding to a chi-square \(\chi^2 = 14.68\) and a reduced chi-square \(\chi^2_\nu = 0.51\). In the present study, six supervised machine learning regression algorithms were employed to reconstruct the Hubble parameter function from the training set and evaluate their predictive performance on the test set: Enhanced Linear Regression, Physics-Informed Linear Regression, Enhanced Artificial Neural Network (ANN), Enhanced Support Vector Regression (SVR), Enhanced Random Forest Regression, and Gradient Boosting Regression. Performance metrics used to assess the models include the coefficient of determination \(R^2\), root mean square error (RMSE), chi-square statistics, reduced chi-square, and mean absolute deviation in the parameter \(\alpha\). Among these, Enhanced SVR provided the best performance with a test \(R^2 = 0.8690\), RMSE of 15.80 km/s/Mpc, \(\chi^2 = 14.27\), reduced \(\chi^2_\nu = 0.4921\), and mean absolute deviation \(|\Delta \alpha| = 0.0725\).
Different graphical illustrations of the ML techniques have been presented in figures (\ref{fig12}-\ref{fig16}). These illustrations reveal a strong agreement between the Hubble parameter predictions from the machine learning models, the theoretical RCHDE model, and observational data. Residuals between predicted and observed \(H(z)\) values are centered around zero and lie within expected uncertainty bounds, demonstrating unbiased and robust predictions. Probability distributions of normalized residuals for SVR, ANN, and linear regression confirm their predictive accuracy. Overall, the use of machine learning in this analysis not only accelerates parameter estimation and increases reliability but also enables the extraction of subtle cosmological information from complex data, providing a powerful framework for probing cosmic expansion and dark energy models.

\section{Stability of the model: Sound Speed Squared Analysis}
A vital test of the physical feasibility of cosmological theories is provided by the propagation of perturbations in the DE sector. The RCHDE model's stability against tiny fluctuations is directly measured by the sound speed squared, which is the derivative of pressure with regard to energy density. This variable defines whether the DE fluid evolves in a physically consistent way by encoding how pressure reacts to energy density perturbations. In order to evaluate stability conditions and the general resilience of the scenario, we examine the evolution of the sound speed squared with redshift and model parameters in this section of the RCHDE framework. We also conduct a comparative sound speed squared analysis between RCHDE, RHDE, and RGBHDE to differentiate between the efficiency of the models as far as stability is concerned. The squared speed of sound is defined as (Y. S. Myung 2007)
\begin{equation}
v_{s}^{2}=\frac{dp}{d\rho}=\frac{\dot{p}}{\dot{\rho}}
\end{equation} 
where dot represents derivative with respect to time and $p$ and $\rho$ are pressure and energy density of DE. If $v_{s}^{2}>0$, we have a stable cosmological model with oscillatory perturbations. If $v_{s}^{2}<0$, then we have an unstable model where the perturbations grow exponentially. Using the eqns.(\ref{denrc2}) and (\ref{pde}) we will get $v_{s}^2$ in terms of the Hubble parameter $H$ and its derivatives. But ideally we will need to derive $v_{s}^2$ in terms of the redshift $z$ so that we can get a proper idea of $v_{s}^2$ with the evolution of the universe. So we need to consider an ansatz for $H$. There can be various ways of considering this, but here we will resort to the simplest power law ansatz given by $H=H_{0}\left(1+z\right)^{m}$, where $H_{0}$ is the present value of the Hubble parameter and $m>0$ is a constant (V. Sahni \& A. Starobinsky (2000); D. Jain et al., 2013; E. J. Copeland et al., 2006). The scaling characteristic of the Hubble parameter in FLRW cosmology, where various cosmic components produce power-law evolution, serves as the inspiration for this power-law parametrization. The dominating component's equation of state is efficiently encoded by the parameter $m$ which also offers a practical phenomenological framework for studying cosmic expansion and late-time acceleration. Now with this set-up we get the $v_{s}^2$ for the RCHDE model as
\begin{equation}
v_{s}^{2}=\frac{\left(1+z\right)^{-2m}\left[2H_{0}m^{2}\left(1+z\right)^{m}\chi+3\rho_{0}\left(1+z\right)^{3}\kappa^{2}\chi-3H_{0}^{2}m\left(1+z\right)^{2m}\left\{3\alpha\left(2m-5\right)\chi+2\times 6^{1/3}\left(3m-2\right)\lambda\right\}\right]}{18H_{0}^{2}m\left[\left(2m-5\right)\chi+12H_{0}^{8}\left(3m-2\right)\left(1+z\right)^{8m}\lambda\tilde{\beta}^{2}\right]} 
\end{equation}
where $\chi=\left\{H_{0}^{2}\alpha\left(3m-2\right)\left(1+z\right)^{2m}\right\}^{2/3}$.

\begin{figure}
    \centering
    \includegraphics[width=0.7\linewidth]{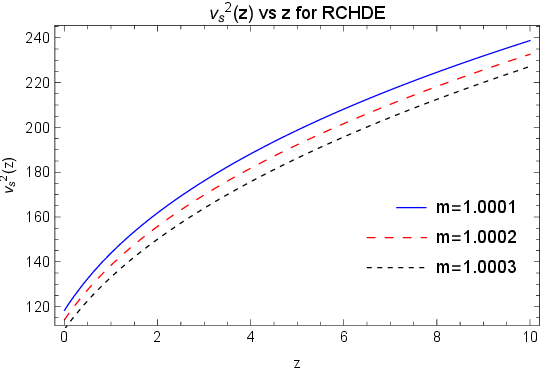}
    \caption{Squared speed of sound against redshift for the RCHDE model for different values of $m$. Here we have taken $\kappa=1$ and other parameters are considered as their best fit values given in table (\ref{tab:parameter_constraints}).}
    \label{figv1}
\end{figure}

\begin{figure}
    \centering
    \includegraphics[width=0.7\linewidth]{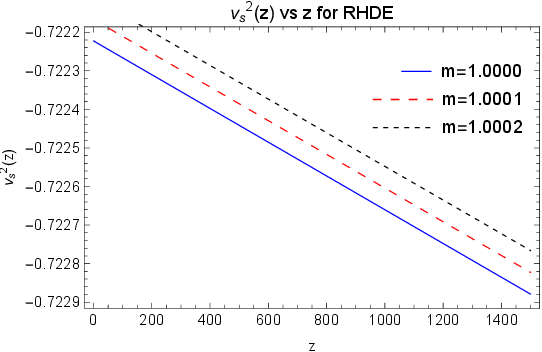}
    \caption{Squared speed of sound against redshift for the RHDE model for different values of $m$. Here we have taken $\kappa=1$ and other parameters are considered as their best fit values.}
    \label{figv2}
\end{figure}

\begin{figure}
    \centering
    \includegraphics[width=0.7\linewidth]{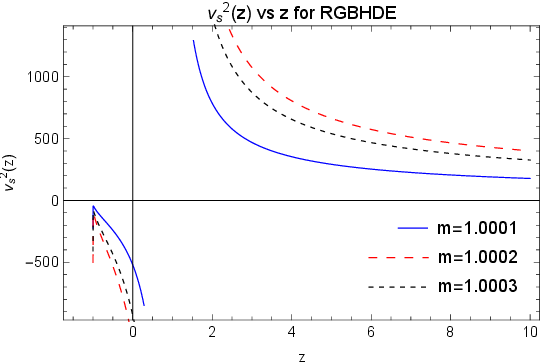}
    \caption{Squared speed of sound against redshift for the RGBHDE model for different values of $m$. Here we have taken $\kappa=1$ and other parameters are considered as their best fit values.}
    \label{figv3}
\end{figure}

In fig.(\ref{figv1}) we have generated the plot of $v_{s}^{2}$ against $z$. It is evident from the plot that the squared speed of sound clearly remains at the positive level throughout the evolution history of the universe. It might seem that the model is stable against density perturbations. But from the figure we see that that trajectory of $v_{s}^2$ starts from around 120 at around $z=0$ and increases further as $z$ increases. This indicates that the model highly suffers from instabilities linked to superluminal propagation. This is a drawback of the model. So as far as stability is concerned, RCHDE does not have any upper hand over RHDE. RHDE suffers from gradient/tachyonic instabilities ($v_{s}^{2}<0$), while RCHDE suffers from superluminal instabilities ($v_{s}^{2}>1$). In figs.(\ref{figv2}) and (\ref{figv3}) we have shown plots of $v_{s}^{2}$ vs $z$ for the RHDE and RGBHDE models respectively. From fig.(\ref{figv2}) we see that the trajectories of $v_{s}^2$ remain at negative levels showing that the RHDE model is classically unstable throughout the evolution history of the universe. From (\ref{figv3}) it is evident that for higher redshifts (early universe) the RGBHDE model is stable, but as we move towards $z=0$ (current epoch) the model suffers from instability. The plot also suggests that this instability is expected to continue in the future times ($z<0$). In a nutshell we see that the RHDE model is classically unstable for almost all redshifts. Including Gauss-Bonnet corrections solves the problem for early universe, but for late times the RGBHDE model still remains unstable, including the current epoch. Finally, when we add the cubic corrections, we see that the model becomes stable for all epoch against the density perturbations. This illustrates enhanced physical viability of the RCHDE model compared to the other models.

\section{Testing the Cosmic Coincidence Problem}
In recent years, the cosmic coincidence problem (CCP) (I. Zlatev et al. 1999; P. J. Steinhardt et al. 1999) has become a significant concern for a number of otherwise successful theories of the universe. The densities of the matter sector, and dark energy sector of the universe are nearly equal in late times, according to recent cosmological observations. When we tie this discovery to the notion that the universe's matter and energy components formed independently from distinct mass scales in the early universe, a difficulty arises. In the late universe, how do they reconcile to the same mass scales? This is a significant cosmological issue that stems from the models themselves. Numerous attempts to address the coincidence problem can be widely found in the literature (Y. Bisabr 2010; P. Rudra 2015(a); P. Rudra 2015(b); P. Rudra 2015(c); P. Huang \& Y-C. Huang (2013)). The ones that consider an appropriate interaction between the universe's matter and energy components are the most impressive. This method takes advantage of the fact that the two sectors of the cosmos have evolved together, interacting with one another, rather than independently from distinct mass scales. This allows matter and energy to move between the two components. A stationary situation is observed in the current universe as a result of this interchange, which dilutes any differences in density.

Here, we will explore the cosmic coincidence problem, to check the efficiency of the model. We define a density ratio parameter
\begin{equation}
r(z)=\frac{\rho_{m}(z)}{\rho_{DE}(z)}
\end{equation}
With this parameter, we can re-frame the CCP as why the value of $r(z)\sim \mathcal{O}(1)$ in the present time. So our work is to determine how fast $r$ evolves and how long it stays near unity for different models. By comparing the scenario for different models we can differentiate between the models and comment on their efficiency to tackle the CCP. To further refine the parameter and to get a more powerful one we define a dimensionless coincidence severity index (CSI) as
\begin{equation}
\mathcal{C}(z)=\left|\frac{d \ln{r}}{d \ln{a}}\right|  
\end{equation}
Here $\ln{a}$ measures cosmic time evolution and $\ln{r}$ removes scale dependence. So this is a very useful construction giving a normalized rate of change of the density ratio. For $\mathcal{C}>>1$ we have rapid evolution, indicating severe coincidence problem. For $\mathcal{C}\sim 1$, we have a moderate evolution, and for $\mathcal{C}<<1$, we have very slow evolution, indicating alleviated coincidence problem. For a non-interacting scenario, the CSI reduces to 
\begin{equation}
\mathcal{C}(z)=3\left|w_{DE}\right|
\end{equation}
For the present time we have 
\begin{equation}
\mathcal{C}_{0}=\left|\frac{d \ln{r}}{d \ln{a}}\right|_{z=0}
\end{equation}
Previously, we have already obtained the expressions for $\rho_{DE}$ and $p_{DE}$ in terms of the Hubble parameter and its time derivative. Now to obtain the above defined coincidence parameters in terms of $z$ we need to consider an ansatz for the background evolution of $H(z)$. There are many redshift parameterization of Hubble parameter and the equation of state parameter available in the literature. In this study, we choose the same power law ansatz for the Hubble parameter given by $H(z)=H_{0}\left(1+z\right)^{m}$, that was used in the previous section. This is a very sensible choice considering that power law forms are known to be perfectly aligned with the cosmological observations and are consistent with the dynamics of the universe. We will also consider an interaction between matter and dark energy given by $Q=3bH\rho_{DE}$, where $b>0$ is the interaction coefficient or dimensionless coupling constant (L. Amendola 2000; W. Zimdahl \& D. Pavon (2001); B. Wang et al., 2016). This interaction term is frequently used in the literature and is dimensionally consistent. This choice suggests a transfer of energy from dark energy to matter, which slows the density ratio's growth and resolves (or alleviates) the cosmic coincidence issue. Including the interaction term in the set-up we get the CSI as
\begin{equation}
\mathcal{C}(z)=3\left|w_{DE}+b\left(1+r\right)\right|
\end{equation}

\begin{figure}
    \centering
    \includegraphics[width=0.7\linewidth]{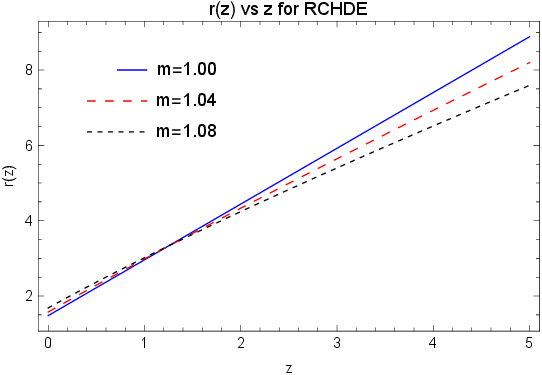}
    \caption{Density ratio parameter $r(z)$ against $z$ for the RCHDE model for different values of $m$. Here we have taken $\kappa=1$ and other parameters are considered as their best fit values.}
    \label{figc1}
\end{figure}

\begin{figure}
    \centering
    \includegraphics[width=0.7\linewidth]{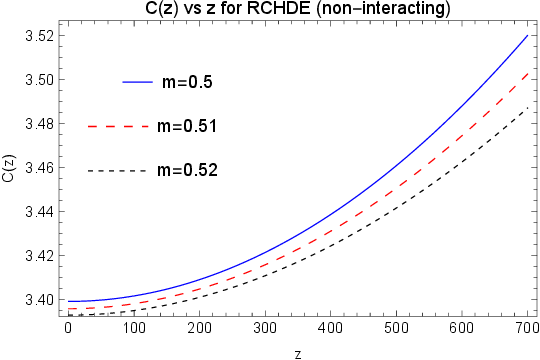}
    \caption{Coincidence severity index $C(z)$ (for the non-interacting case) against $z$ for the RCHDE model for different values of $m$. Here we have taken $\kappa=1$ and other parameters are considered as their best fit values.}
    \label{figc2}
\end{figure}

\begin{figure}
    \centering
    \includegraphics[width=0.7\linewidth]{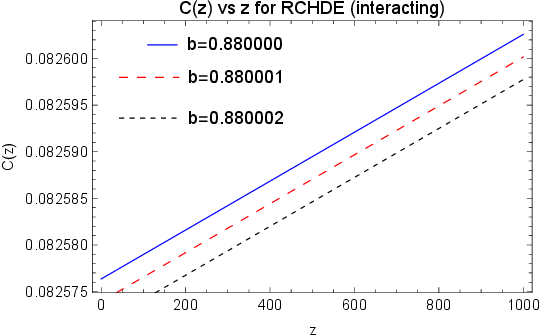}
    \caption{Coincidence severity index $C(z)$ (for the interacting case) against $z$ for the RCHDE model for different values of the interaction coefficient $b$. Here we have taken $\kappa=1$ and other parameters are considered as their best fit values.}
    \label{figc3}
\end{figure}

\begin{figure}
    \centering
    \includegraphics[width=0.7\linewidth]{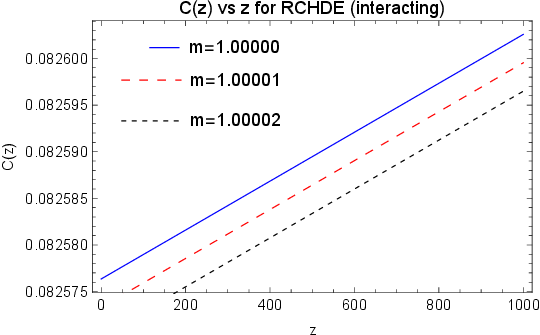}
    \caption{Coincidence severity index $C(z)$ (for the interacting case) against $z$ for the RCHDE model for different values of $m$. Here we have taken $\kappa=1$ and other parameters are considered as their best fit values.}
    \label{figc4}
\end{figure}

We have plotted the coincidence parameters for the RCHDE model in figs.(\ref{figc1}), (\ref{figc2}), (\ref{figc3}) and (\ref{figc4}). From fig.(\ref{figc1}) we see that the density ratio parameter $r(z)\sim O(1)$ in the present time ($z=0$). This shows that $r$ evolves slowly and reaches a stable fixed point. In fig.(\ref{figc2}) we see that the coincidence severity index $C(z)>1$, indicating that in the absence of interaction, there is no solution to the CCP. In fig.(\ref{figc3}) we see for an interacting scenario between DE and matter $C(z)<<1$ indicating an alleviated CCP. Moreover, we see that for greater interaction, we have greater alleviation of CCP as expected. In fig.(\ref{figc4}) we have a similar plot for $C(z)$ showing the dependency on $m$. We see that there is a significant alleviation of CCP, which becomes better with the increase of $m$.

\begin{figure}
    \centering
    \includegraphics[width=0.7\linewidth]{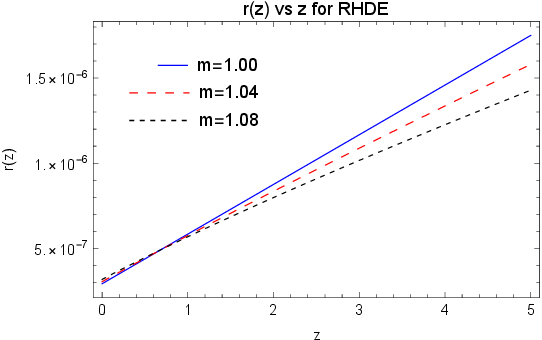}
    \caption{Density ratio parameter $r(z)$ against $z$ for the RHDE model for different values of $m$. Here we have taken $\kappa=1$ and other parameters are considered as their best fit values.}
    \label{figc5}
\end{figure}

\begin{figure}
    \centering
    \includegraphics[width=0.7\linewidth]{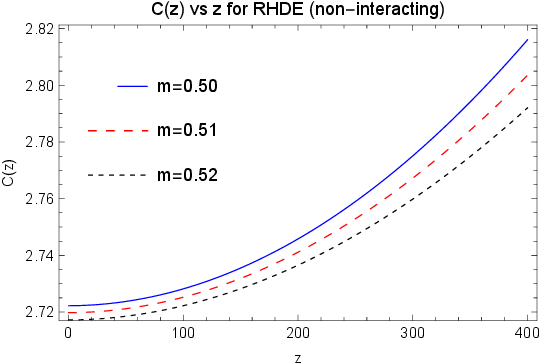}
    \caption{Coincidence severity index $C(z)$ (for the non-interacting case) against $z$ for the RHDE model for different values of $m$. Here we have taken $\kappa=1$ and other parameters are considered as their best fit values.}
    \label{figc6}
\end{figure}

\begin{figure}
    \centering
    \includegraphics[width=0.7\linewidth]{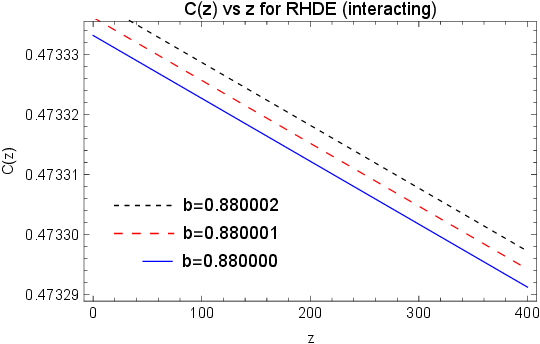}
    \caption{Coincidence severity index $C(z)$ (for the interacting case) against $z$ for the RHDE model for different values of the interaction coefficient $b$. Here we have taken $\kappa=1$ and other parameters are considered as their best fit values.}
    \label{figc7}
\end{figure}

\begin{figure}
    \centering
    \includegraphics[width=0.7\linewidth]{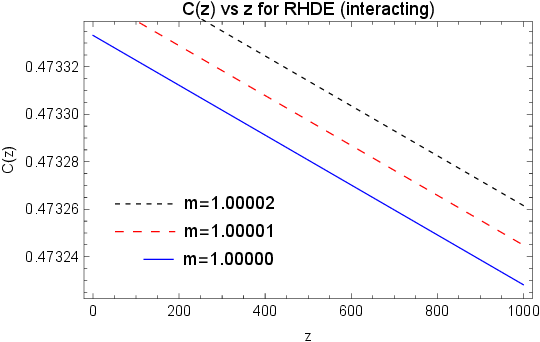}
    \caption{Coincidence severity index $C(z)$ (for the interacting case) against $z$ for the RHDE model for different values of $m$. Here we have taken $\kappa=1$ and other parameters are considered as their best fit values.}
    \label{figc8}
\end{figure}

\begin{figure}
    \centering
    \includegraphics[width=0.7\linewidth]{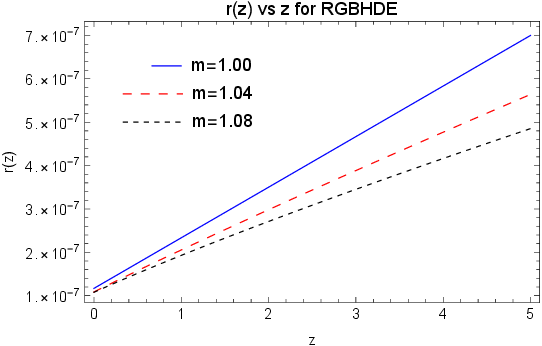}
    \caption{Density ratio parameter $r(z)$ against $z$ for the RGBHDE model for different values of $m$. Here we have taken $\kappa=1$ and other parameters are considered as their best fit values.}
    \label{figc9}
\end{figure}

\begin{figure}
    \centering
    \includegraphics[width=0.7\linewidth]{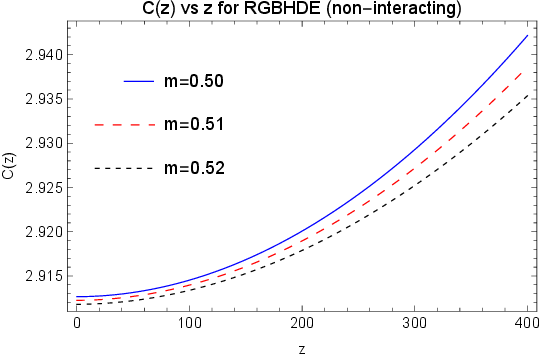}
    \caption{Coincidence severity index $C(z)$ (for the non-interacting case) against $z$ for the RGBHDE model for different values of $m$. Here we have taken $\kappa=1$ and other parameters are considered as their best fit values.}
    \label{figc10}
\end{figure}

\begin{figure}
    \centering
    \includegraphics[width=0.7\linewidth]{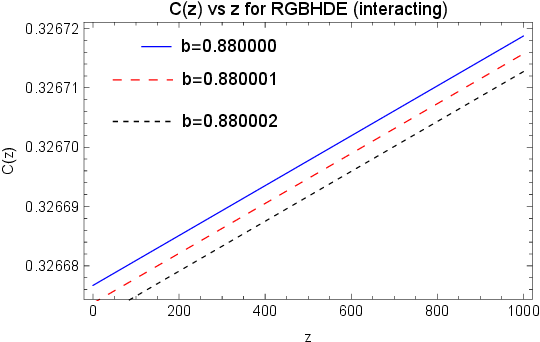}
    \caption{Coincidence severity index $C(z)$ (for the interacting case) against $z$ for the RGBHDE model for different values of the interaction coefficient $b$. Here we have taken $\kappa=1$ and other parameters are considered as their best fit values.}
    \label{figc11}
\end{figure}

\begin{figure}
    \centering
    \includegraphics[width=0.7\linewidth]{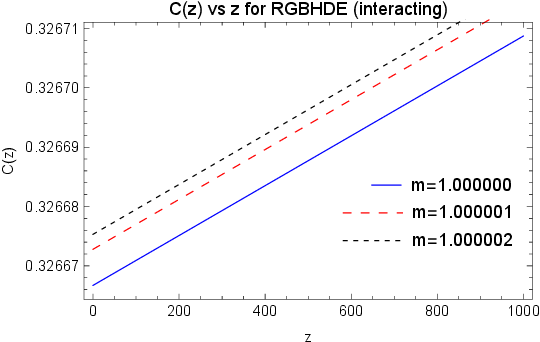}
    \caption{Coincidence severity index $C(z)$ (for the interacting case) against $z$ for the RGBHDE model for different values of $m$. Here we have taken $\kappa=1$ and other parameters are considered as their best fit values.}
    \label{figc12}
\end{figure}

We have also performed the coincidence parameter analysis for the RHDE and RGBHDE model, to differentiate the models and identify their efficiency in a comparative scenario.  We have plotted the coincidence parameters for the RHDE model in figs.(\ref{figc5}), (\ref{figc6}), (\ref{figc7}) and (\ref{figc8}). From fig.(\ref{figc5}) we see that the density ratio parameter $r(z)\sim O(10^{-7})$ in the present time ($z=0$). This shows that matter decays very fast (or DE dominates too early) and we do not get a comparable scenario between matter and DE as the observations suggest. In fig.(\ref{figc6}) we see that the coincidence severity index $C(z)>1$, indicating that in the absence of interaction, there is no solution to the CCP. This case is similar to that of RCHDE model. In fig.(\ref{figc7}) we see for an interacting scenario between DE and matter $C(z)<1$ indicating only a mild alleviation to CCP. Moreover, we see that for greater interaction, there is no sign of greater alleviation of CCP. Moreover, as we approach the present time from larger redshifts we see that there is an increase in the value of $C(z)$ indicating an increased CCP. A similar situation is seen in fig.(\ref{figc8}) where the plot shows the dependence of $C(z)$ on $m$. In the figs.(\ref{figc9}), (\ref{figc10}), (\ref{figc11}) and (\ref{figc12}) coincidence parameters are plotted for the RGBHDE model. From fig.(\ref{figc9}) we see that the density ratio parameter $r(z)\sim O(10^{-7})$ in the present time ($z=0$), similar to the RHDE model. This shows that matter decays very fast and we do not get a comparable scenario between matter and DE as the observations indicate. In fig.(\ref{figc10}) we see that the coincidence severity index $C(z)>1$, indicating that in the absence of interaction, there is no solution to the CCP. This case is similar for all the three models. In fig.(\ref{figc11}) we see for an interacting scenario between DE and matter $C(z)<1$ indicating only a mild alleviation to CCP, just like RHDE. But here we see that for greater interaction, there are signs of greater alleviation of CCP, unlike the RHDE case. Moreover, as we approach the present time from larger redshifts we see that there is a decrease in the value of $C(z)$ indicating alleviation of CCP. This is a better scenario as compared to RHDE, but not as good as RCHDE. A similar scenario s observed in fig.(\ref{figc12}) where the dependence of of $C(z)$ on $m$ is portrayed. In a nutshell, the RHDE model provides only a mild alleviation of the CCP. The scenario improves for the RGBHDE model, but not good enough to be considered as a significant solution to CCP. The RCHDE model produces a smoother dynamical evolution of the density ratio and extends the epoch where matter and dark energy remain comparable, thereby offering the strongest alleviation of the CCP among the three models. It must be stated here that the above analysis is obviously dependent on the parameter space, but since we have used the observationally constrained values (determined in section 3) it adds significant value to the analysis.

\section{Conclusion}
In this work, we have explored the Ricci-cubic holographic dark energy model formed from the cubic invariant, which in turn is formed from the cubic contractions of the Riemann and Ricci tensors. We have performed a detailed observational data analysis of the cosmological model using different data sets like the Hubble data, cosmic-chronometer data, Baryon-acoustic oscillation data, and also data from gamma-ray bursts. We have also explored a combination of datasets for better results. The objective was to constrain the model by finding best-fit estimations of the free parameters of the model. To do this, we have used the Markov chain Monte-Carlo sampling technique. The constraints on the model parameters are obtained via Bayesian inference. Contour plots have been obtained for the model parameters, showing their marginalized and joint probability distributions. The plots show the different confidence intervals where these parameters may lie, thus constraining them considerably. We have also compared our constrained model with the $\Lambda$CDM model to verify and validate our work and found remarkable agreement. To do this, the best-fit regression lines are found for the constrained model and compared with the standard $\Lambda$CDM model. In some instances, our model showed superior results compared to the $\Lambda$CDM model, thus indicating the importance of our model. Apart from this, we have checked the Hubble tension for our model. The best-fit value obtained in our model exhibits a moderate tension of approximately $2.3\sigma$ with respect to the reference value for $\Lambda$CDM. While this indicates a noticeable discrepancy, it remains significantly lower than the $\sim 5\sigma$ tension typically reported between early- and late-Universe measurements, suggesting a partial alleviation of the Hubble tension.

To complement this data analysis mechanism, we have also performed an enhanced machine learning analysis using observational Hubble parameter data. This approach serves to validate the model’s predictive power through independent, data-driven regression techniques. The use of machine learning in this analysis not only accelerates parameter estimation and increases reliability but also enables the extraction of subtle cosmological information from complex data, providing a powerful framework for probing cosmic expansion and dark energy models. Different graphical illustrations of the machine learning techniques have been presented to understand the results. These illustrations reveal a strong agreement between the Hubble parameter predictions from the machine learning models, the theoretical RCHDE model, and observational data. 

Using higher curvature invariants makes the model more naturally connected with the geometry of spacetime in general relativity and its extensions. This improves the purely phenomenological nature of RHDE. In four dimensions, the Gauss-Bonnet invariant is mostly topological, but cubic invariants are not. So they can influence cosmic dynamics more naturally and influence Einstein’s field equations directly. This solves the issue with the Gauss-Bonnet term of RGBHDE. This is a very important point to note for our model. Our model has more parameters compared to RHDE and RGBHDE, and so it presents much richer cosmological dynamics. However, some cubic invariant combinations can produce positive sound speed squared, potentially avoiding the gradient instability problem found in RHDE. Our model is expected to substantially alleviate the cosmic coincidence problem. This is because the cubic model contains terms like $H^6$, $H^4 \dot{H}$, $H^{2}\dot{H}^2$. Due to the presence of these terms, the dark energy density can naturally track the evolution of matter density. In this work, we have substantially improved the Ricci-cubic holographic dark energy model and rendered it useful for any cosmological analysis. Our work has shown that this model can be considered as a comprehensive candidate of dark energy that can explain the late-time universe. We believe that this work is a substantial improvement and addition to the existing literature on this topic, especially considering the advanced computer algorithms that have been used. Finally to check the efficiency of our model, in comparison to other holographic dark energy models like the Ricci holographic dark energy, Ricci-Gauss-Bonnet holographic dark energy, we have done a stability analysis using the squared speed of sound and also tested the cosmic coincidence problem for the models. It was seen that the RCHDE model did not not suffer from the gradient instability ($v_{s}^{2}<0$) like the RHDE model, but it suffers from superluminal instability ($v_{s}^{2}>1$). So the RCHDE model does not hold any upper hand over its counterparts as far as stability is concerned. In the cosmic coincidence problem analysis, RCHDE model produced a significant alleviation to the coincidence problem, outshining its counterparts. These two analysis show that the RCHDE model has some edge over its counterparts, and solves important problems that plagues other holographic dark energy models. Currently we do not make tall claims regarding the success of our RCHDE model. But at least with the analysis that is performed in this paper, it seems to show positive signs and satisfactory results. The model demands our attention and we need to do more investigations on the model.

%%%%%%%%%%%%%%%%%%%%%%%%%%
\section*{Acknowledgments}
%%%%%%%%%%%%%%%%%%%%%%%%%%
PR acknowledges the Inter-University Centre for Astronomy and Astrophysics (IUCAA), Pune, India for granting a visiting associateship. The authors also thank the anonymous referee for his/her invaluable comments that helped them to improve the quality of the manuscript.

\section*{Data Availability Statement}
The data used in the paper are widely available in the literature. The references from which the data have been taken are cited in the paper in the appropriate places.

\section*{Conflict of Interest}

There are no conflicts of interest.

\section*{Funding Statement}

There is no funding to report for this article.
%%%%%%%%%%%%%%%%%%%%%%%%%%%%%

\end{document}